# Real time observation of cuprates structural dynamics by Ultrafast Electron Crystallography

Authors: F. Carbone[1,2*], N. Gedik[1,3], J. Lorenzana[4], and A.H. Zewail[1]

[1]*Physical Biology Center for Ultrafast Science and Technology, Arthur Amos Noyes Laboratory of Chemical Physics, California Institute of Technology, Pasadena, CA 91125, USA.*
[2] *Current address: Laboratory of Ultrafast Spectroscopy, Ecole polytechnique fédérale de Lausanne, 1015 Lausanne, CH.*
[3] *Current address: MIT Department of Physics 77 Mass. Ave., Bldg. 13-2114 Cambridge, MA 02139 USA.*
[4] *Universita` di Roma La Sapienza, Piazzale Aldo Moro 2, 00185 Roma, Italy.*

*Authors to whom correspondence should be addressed. E-mail:

fabrizio.carbone@epfl.ch

# ABSTRACT

The phonon-mediated attractive interaction between carriers leads to the Cooper pair formation in conventional superconductors. Despite decades of research, the glue holding Cooper pairs in high-temperature superconducting cuprates is still controversial, and the same is true as for the relative involvement of structural and electronic degrees of freedom. Ultrafast electron crystallography (UEC) offers, through observation of spatio-temporally resolved diffraction, the means for determining structural dynamics and the possible role of electron-lattice interaction. A polarized femtosecond (fs) laser pulse excites the charge carriers, which relax through electron-electron and electron-phonon coupling, and the consequential structural distortion is followed diffracting fs electron pulses. In this review, the recent findings obtained on cuprates are summarized. In particular, we discuss the strength and symmetry of the directional electron-phonon coupling in $Bi_2Sr_2CaCu_2O_{8+\delta}$ (BSCCO), as well as the *c*-axis structural instability induced by near-infrared pulses in $La_2CuO_4$ (LCO).

The theoretical implications of these results are discussed with focus on the possibility of charge stripes being significant in accounting for the polarization anisotropy of BSCCO, and cohesion energy (Madelung) calculations being descriptive of the *c*-axis instability in LCO.

# I INTRODUCTION

Despite two decades of intense research, the mechanism of high temperature superconductivity in cuprates is still unclear (*1*). Besides their high temperature superconductivity, cuprates display a rich, yet poorly understood phase diagram covering electronic and structural phase transitions as a function of temperature, chemical doping and magnetic field strength (*2*). The unique behavior seen in these materials is a result of the delicate interplay between charge, spin and lattice excitations. As a result of these couplings between different degrees of freedom, there usually exist several competing states, which give rise to multiple phases (*3*). Understanding the dynamics between different electronic and structural phases of these materials is significant for an eventual understanding of superconductivity.

Several experimental techniques have been applied to the study of these materials. Information on structural dynamics has been obtained through Raman spectroscopy (*4*), X-ray absorption spectroscopy (XAS) (*5*), Angle resolved photoemission spectroscopy (ARPES) (*6-8*) (*9*), and oxygen isotope substitution studies (*10-13*). In parallel, the effect of strong electron-electron interactions have been revealed by optical spectroscopy (*14, 15*), inelastic neutron scattering (*16*), ARPES (*17*), transport (*18*), and scanning tunneling microscopy (*19*). Theoretically, a plethora of exotic many-body entities can emerge, either from electron-electron correlations (*20*), or electron-lattice interactions (*21, 22*), but to date, a consensus has not been reached yet on the exact nature of the ground state of cuprates superconductors.

The time scale for electron-electron interactions is usually much faster than that of lattice dynamics. As a result, the crystal structure can be considered at equilibrium while electronic scattering phenomena are taking place. However, this scenario can be perturbed when strong electron-phonon coupling (larger than what is found in $MgB_2$, which has a $T_c$ close to 40 K (*23*)), and strong electron-electron correlations are involved. When the electron-phonon coupling time reaches the tens of fs scale, it becomes comparable to the time scale for inter-electronic scatterings and magnetic interactions (*24*). In this situation, the conventional approximations need to be revised (*25*). Such issue of time scales can be addressed by UEC by resolving the atomic motions in real time.

Optical time-resolved techniques have been extensively used to study cuprates (*26-31*). The most commonly involved to date is the pump-probe spectroscopy method. This technique is sensitive to the dynamics of electrons, and it is based on measuring the photoinduced changes in the reflectivity or transmission of an optical probe pulse in response to an absorption by a strong pump pulse. In the superconducting state, fs pulses were used to break Cooper pairs and re-pairing dynamics of the resulting quasiparticles were studied across the phase diagram. By changing the wavelength of the probe pulse from terahertz (*32*) to mid-infrared (*28*), and to optical frequencies (*26, 29*), dynamics of the superconducting condensate or quasiparticle sub-system could be studied as a function of time. These experiments yielded both the elastic and inelastic quasiparticle scattering rate (*33*), the quasiparticle diffusion rate (*34*) as a function of temperature, and the excitation density (and doping) in cuprates (*29*).

All of the previous time-resolved studies in cuprates were based on the probing of dynamics, as reflected in electrons response in spectroscopic probing, but with no direct information about the structural dynamics of the underlying lattice. In the superconducting state, for example, the quasiparticles give their energy to a boson when they recombine to make Cooper pairs, and optical experiments cannot directly follow the evolution of the system after this point, since the resulting excitation does not cause a significant reflectivity change. Whether the electrons are directly coupled to phonons or to other collective excitations is still an open question. The energy is ultimately transferred to the lattice as heat, but to date no direct time-resolved measurement of this process was made, nor do we know the actual structural deformations especially in relation to symmetry and direction of electron-phonon interactions.

In this review, we describe recent results obtained by UEC. The temporal evolution of the crystal structure of BSCCO samples, following polarized carrier excitation by a fs pulse, for different temperatures (for the metallic and superconducting states) and doping levels (from underdoped to optimally doped) has been reported (*24*). Specifically, different compositions were investigated, by varying the doping level and number of Cu–O planes per unit cell in the BSCCO family, and by probing optimally doped LCO (La, Cu, O) nano-islands. In these experiments, the initial fs excitation drives the system from the superconducting into the metallic phase (*30*), breaking Cooper pairs (*29*). With the electron and lattice temperatures being vastly different (see below), energy of carriers is lowered through electron-phonon coupling, and structural distortions are observable in diffraction (*24, 35*).

By varying the polarization of carrier excitation in BSCCO, major differences in the decay of Bragg diffraction were observed, for the *c*-axis structural dynamics. The striking polarization effect for this *c*-axis motion is consistent with a highly anisotropic electron-phonon coupling to the $B_{1g}$ out-of-plane buckling mode (50 meV), with the maximum amplitude of atomic motions being ~0.15 Å. Out-of-plane motions give also rise to an out-of-equilibrium, structural phase transition in LCO nano-islands. For the latter, the *c*-axis lattice parameter value was, in fact, found to jump between two structures isobestically, in response to the optical excitation of the charge carriers (*35*).

In the following, we will describe these experimental findings in details, and discuss the theoretical implications of the results. In particular, we will discuss the role of polarized light excitation in cuprates and the possibility that charge stripes may be behind the observed anisotropy in BSCCO. We will also show that simple cohesion energy calculations can account for the observed *c*-axis instability in LCO, based on the assumption that infrared pulses photo-dope the Cu-O planes through a charge transfer process.

**II EXPERIMENTAL**

**II. a. The UEC apparatus**

The experimental setup, displayed in Fig. 1, consists of a fs laser system and three connected ultrahigh-vacuum chambers (for diffraction, load lock, and sample preparation/characterization). The laser system generates amplified pulses that are centered at 800 nm (1.55 eV), with a pulse-width of 120 femtosecond and energy of 1.8 mJ per pulse at a repetition rate of 1 kHz. A beam splitter was placed to separate the beam into two arms: a pump beam used to excite the sample and a second beam for generating the probing electron packets. The second beam was obtained by tripling (266 nm, 4.65 eV), via third harmonic generation in a nonlinear optical device, the fundamental frequency. The UV pulses impinged on a back-illuminated photocathode thus generating the fs electron bunches through the photoemission process. The time delay between the excitation and probing electron pulses was changed by adjusting the relative optical path length between the two arms using a motorized delay line.

The probing electrons were accelerated up to 30 keV, giving a de Broglie wavelength of ~0.07 Å; they were focused by a magnetic lens, and directed to the sample with an incidence angle typically between $0^0$ and $4^0$. The light excitation pulse was focused onto the sample in a cylindrical spot, in order to overlap with the probe electrons footprint. The temporal mismatch, due to the difference in velocity between photons and electrons, was suppressed by tilting the optical wavefront of the laser pulses (*36*). With this arrangement, the electrons and photons overlapped in time, with no delay, across the entire area probed (as pictorially displayed in Fig. 1). The size of the laser spot was carefully measured to be about 430 μm by 3 mm FWHM by using a separate CCD camera. The electron beam had a cross section diameter of 200 μm FWHM, as measured on the screen; in these studies, it contained ~1000 electrons per pulse having a duration between 0.5 to 1 ps. The resulting average current density of electrons was relatively small, on the order of 0.1 pA/mm$^2$, which is not sufficient to induce damage.

The samples were mounted on a high precision five-axis goniometer capable of providing rotations with angular resolution of $0.005^0$. An external cryostat was coupled to the sample holder through a flexible copper braid, to vary the sample temperature between 10 and 400 K. The diffraction patterns were recorded using a low-noise image-intensified CCD camera assembly capable of single-electron detection. Typically, the averaging time for a single diffraction frame was around 10 seconds. Several diffraction frames were averaged over multiple time scans in order to obtain a good signal-to-noise ratio. The data were processed with home-built computer interface.

### II. b. BSCCO samples and static diffraction

The optimally-doped samples of Bi2212 and Bi2223 were grown by the "travel solvent floating zone" technique, described in ref. (*37*). The superconducting transition temperature was found to be $T_c$ = 91 K in Bi2212 ($\Delta T_c$ = 1 K), and $T_c$ = 111 K in Bi2223 ($\Delta T_c$ = 4 K). The underdoped Bi2212 sample was grown by the self-flux method (*38*), annealed in an oxygen-deficient atmosphere, and its transition temperature was found to be $T_c$ = 56 K ($\Delta T_c$ < 6 K). The magnetic susceptibility curves for two representative samples are given in Ref. (*24*). All samples were cleaved *in situ* at low temperature (20 K) prior to diffraction experimental studies in order to ensure the presence of a high-quality, clean surface.

In Fig. 2, we display the static diffraction pattern obtained from the optimally doped Bi2212 sample. The patterns were recorded in the reflection geometry with the electron beam directed along three different axes, namely the [010], [110], and [100] directions, as displayed in panels A to C; The penetration depth of 30 kV electrons in BSCCO is expected to be less than 100 Å; the *c*-axis lattice parameter is 30 Å, therefore only few unit cells (3-4) are probed by electrons. As a result, the diffraction patterns present rods instead of spots. In Fig. 2 A the (20) rod shows a weak modulation consistent with the *c*-axis value of the material. The diffraction was indexed for the tetragonal structure, giving the in-plane lattice parameters of $a = b = 5.40$ Å, and $c = 30$ Å, consistent with X-ray values. The lattice modulation is resolved along the $b$-axis with a period of 27 Å, again in agreement with the X-ray data (*39*). The in-plane lattice constants, as well as the modulation, were confirmed for the specimens studied using our electron microscope. One micrograph is shown in Fig. 2 D.

In order to quantify the diffraction, different "cuts" along the momentum transfer vector were made; see Fig. 3. The colored arrows in Fig. 2 and 3 indicate the direction in which the cut is performed. From the 2-D data we can extract with precision the in-plane and out-of plane lattice parameters, and the presence of several higher order diffraction features testifies for the good quality of the samples. In Fig. 4, the diffraction image and the 2-D cut parallel to the (11) direction for underdoped Bi2212 (panel A and B) and optimally doped Bi2223 (panel C and D) are also shown. The in-plane lattice parameter of Bi2223 is found to be $a = b = 5.42$ Å, in agreement with earlier X-ray data (*39*).

In Fig. 5, we present the unit cell of Bi2212 (panel A); for clarity the different directions indexed in the diffraction patterns are indicated here. In the same graph, the red arrows departing from the oxygen ions represent the distortion induced by two particular phonon modes (the in-plane breathing and the out-of plane buckling) which will be relevant for the following discussion. In panel B, one can see the effect of the $c$-axis modulation along the $b$-axis of a Bi2223 crystal. The modulation is present in all Pb-free BSCCO samples, and is responsible for the satellites observed in the diffraction pattern recorded with electrons probing parallel to $b$ (Fig. 2 C), and in transmission (Fig. 2 D).

**II. c. LCO sample and static diffraction**

The La$_2$CuO$_{4+\delta}$ (LCO) film used was 52 nm thick (*35*); it has been grown on LaSrAlO$_4$ (LSAO) using a unique atomic-layer molecular beam epitaxy (MBE) (*40*) system equipped with 16 metal sources (thermal effusion cells), a distilled ozone source, and a sophisticated, real-time, 16-channel rate monitoring system based on atomic absorption spectroscopy. It is also provided with a dual-deflection reflection high-energy electron diffraction (RHEED) system and a time-of-flight ion scattering and recoil spectroscopy (TOF-ISARS) system for real-time chemical analysis of the film surface. These advanced surface-science tools provide information about the film surface morphology, chemical composition, and crystal structure. The films under study were characterized by resistivity, X-ray diffraction (XRD), atomic force microscopy (AFM) and electrostatic force microscopy (EFM). The resistivity showed the onset of superconductivity around 32 K and the X-ray diffraction analysis confirmed the good crystallinity of the film with lattice parameters of $a = b = 3.755$ Å and $c = 13.2$ Å.

The growth of the samples was monitored with RHEED in the MBE chamber. During the growth, the pattern showed sharp streaks consistent with an atomically smooth surface. After the sample was taken out of the growth chamber and transported between laboratories, we could only observe transmission like electron diffraction patterns on top of a broad background intensity, indicating the modification of the original surface and existence of three dimensional structures on the film. Electron diffraction from these structures matches with the structure of the LCO film, as will be shown below. AFM measurements taken on the film after exposure to air showed atomically smooth surfaces (rms roughness in 0.3-0.6 nm range) except for some rare precipitates with the typical width of ca 50-200 nm. AFM topography images show that these precipitates typically have a cylindrical shape with diameters around 50 nm and typical height of 20 nm.

In Fig. 6, shown are static electron diffraction patterns obtained from two different orientations of the sample. In Fig. 6 A, the electron beam is incident at $45^0$ with respect to the in-plane Cu-O bond direction (nodal direction), whereas in Fig. 6 B the beam is incident along the Cu-O bond direction (antinodal direction). The angle of incidence was around $1.5^0$ in both cases. The diffraction patterns observed in both cases are consistent with the LCO crystal structure. We have indexed these patterns based on the tetragonal structure. The obtained lattice constants ($a = b = 3.76$ Å and $c = 13.1\pm0.1$ Å) are in agreement with the aforementioned X-ray diffraction measurements we made on the same film ($a = b = 3.755$ Å and $c = 13.20$ Å). The uncertainty in the lattice constants obtained with electron diffraction comes mainly from the error in the determination of the sample to camera distance. The relative changes in the lattice constants can be measured with much better accuracy (below $\pm0.01$ Å).

The structure of LCO at high temperature is known to be tetragonal (HTT) with space group I4/mmm. Once the sample is cooled down, tilting of the $CuO_6$ octahedra occurs and transition to a low temperature orthorhombic (LTO) phase takes place (*41*). In the undoped compound, this transition occurs at ~530 K. Depending on the oxygen concentration, these tilts can be ordered having a space group of $B_{mab}$ or disordered with a space group of $F_{mmm}$. The tilting of the $CuO_6$ octahedra results in the appearance of weak satellite peaks in the diffraction pattern at locations that are not allowed in the tetragonal symmetry, however these satellite peaks were too weak to be see in our diffraction patterns. Locations of the main lattice Bragg peaks are not affected. For simplicity, we used the tetragonal phase (I4/mmm) for indexing of patterns although the actual space group might not be strictly tetragonal.

## III RESULTS AND DISCUSSION

### III a. The Debye-Waller effect in BSCCO

We begin by discussing the results obtained for BSCCO samples. The temporal evolution of diffraction frames (with polarized excitation) is sensitive to motions of atoms during the structural change. In Fig. 7 A, the intensity decay due to motions of the ions (Debye-Waller effect) of the (00) rod is plotted for three different polarizations ($\vec{E}$) of the excitation pulse: $\vec{E}$ //[010], the direction of Cu–O bonds; $\vec{E}$ //[110], the direction at 45°; and the one at 22°. The data were taken at $T$ = 50 K on an optimally doped Bi2212 sample. At longer times, up to 1 ns, these transients recover very slowly; because of the poor $c$-axis conductivity and metallic $ab$-plane, heat transport is mainly lateral, but is complete on the time scale of our pulse repetition time (1 ms). In Fig. 7 B, another set of data was obtained by rotating the same sample while keeping the polarization parallel to the electron beam direction. The temporal evolution of the (00) diffraction intensity obtained from the two different orientations (electron beam parallel to the Cu–O bond, see diffraction pattern in Fig. 2 A and the corresponding Bragg peak in Fig. 3 A, and at 45°, pattern in Fig. 2 B and corresponding Bragg peak in Fig. 3 B) shows the same anisotropic behavior as that obtained by rotating the polarization, ruling out possible experimental artifacts.

The intensity decay for different polarizations was found to have distinct time constants (see below): the decay is faster when the polarization is along the Cu–O bond and slows down when polarization is along the [110] direction (45° from the Cu–O bond). Such an effect was observed to be even stronger in an underdoped sample. In Fig. 7 C we display the results obtained for underdoped Bi2212 ($T_c$ = 56 K), also at two temperatures. The anisotropy is evident at low temperature, with the Debye-Waller decay being faster again for $\vec{E}$ //[010]. However, at higher temperature, the decay of both polarizations is similar and reaches the fastest profile recorded. Surprisingly, in optimally doped Bi2223, we observed no significant anisotropy even in the low temperature regime (Fig. 7 D). In fact, the intensity decay of the (00) rod for light polarized along [110] becomes essentially that of the [010] direction.

### III b. The electron-phonon coupling parameter

When charge carriers are excited impulsively through light in a crystal, the electron and lattice temperatures are driven out of equilibrium, but they equilibrate through electron-phonon coupling. Excitation of phonons causes the diffraction intensity to change with time, and this decrease mirrors an increase of the mean atomic displacement in the corresponding direction, with a temperature assigned to the displacement through a time-dependent Debye-Waller factor:

$$\ln[I(t)/I_0] = -2W(t) = -s^2 \langle \delta u^2(t) \rangle / 3, \tag{1}$$

where $I(t)$ is the intensity of rod diffraction at a given time $t$ after excitation, $I_0$ is the intensity before excitation, $s$ is the scattering vector, and $\langle \delta u^2(t) \rangle$ is the mean-square atomic displacement. From the results reported here for $[I(t)/I_0]_{min}$, the root-mean-square value for the amplitude of the motion is obtained to be ~0.15 Å for 20 mJ/cm$^2$ fluence.

Given the *c*-axis distance of 30 Å, this represents a change of 0.5% of the *c*-axis; the Cu–O planes instead, separate by 3.2 Å.

We verified that different diffraction orders show changes which scale with the scattering vector, confirming that the observed changes in the diffraction intensity originate from phonon-induced structural dynamics. In a time-resolved diffraction experiment, different Bragg spots at a given time should exhibit intensity changes in accord with the value of the scattering vectors *s* (see Eq. **1**). Therefore, two distinct Bragg diffraction features appearing at $s = s_1$ and $s_2$ should obey the following scaling relation:

$$\frac{\ln(I_{s_1}/I_0)}{\ln(I_{s_2}/I_0)} = (\frac{s_1}{s_2})^2, \qquad (2)$$

In Fig. 8, we plot the intensity changes for two different Bragg spots, recorded in the same pattern, but for different scattering vectors. The apparent scaling confirms that the observed intensity changes are indeed originating from structural motions.

The observed anisotropy of decays with polarization reflects the distinct *c*-axis distortion and the difference in electron-phonon coupling. In order to obtain the magnitude of the couplings we invoked the well-known model of electrons and lattice temperatures, and dividing the lattice modes into those which are strongly coupled to the electrons and the rest which are not (*24*). Thus, the decrease of the intensity at a given time tracks the change of $<\delta u^2(t)>$ with a corresponding effective temperature. For a Debye solid, the atomic displacement can be expressed as

$$\left\langle \delta u^2(t) \right\rangle = \frac{9\hbar^2 \Delta T(t)}{M k_B \Theta_D^2}, \qquad (3)$$

where $M$ is the average mass in the unit cell, $k_B$ is the Boltzmann constant, $\hbar$ is the reduced Planck constant, and $\Theta_D$ is the Debye temperature of the material (*30*). Traditionally, the two-temperature model (*42*) is invoked to describe the laser-induced heating of electrons and phonons, as subsystems, in an elementary metal. Its success is the result of the isotropic electron-phonon coupling in a simple lattice structure, i.e., one atom per primitive unit cell. In complex, strongly correlated materials like high-$T_c$ superconductors, however, such model becomes inappropriate because photoexcited carriers may anisotropically and preferentially couple to certain optical phonon modes, making meaningless the assignment of a single temperature to the whole lattice structure (*30, 43*).

In the three-temperature model described in ref. (*30*), in addition to the electron temperature $T_e$, two temperatures are defined for the lattice part: the hot-phonon temperature, $T_p$, for the subset of phonon modes to which the laser-excited conduction-band carriers transfer their excess energy, and the lattice temperature, $T_l$, for the rest of the phonon modes which are thermalized through anharmonic couplings. As an approximation, the spectrum of the hot phonons $F(\Omega)$ is assumed to follow an Einstein model: $F(\Omega) = \delta(\Omega - \Omega_0)$, where $\delta$ denotes the Dirac delta function, with $\Omega$ being the energy and $\Omega_0$ the energy of a hot phonon. Effectiveness of the energy transfer between the carriers and hot phonons is described by the dimensionless parameter $\lambda$: $\lambda = 2 \int \Omega^{-1} \alpha^2 F d\Omega$, where $\alpha^2 F$ is the Eliashberg coupling function (*42*). The rate equations describing the temporal evolution of the three temperatures are given by:

$$\frac{dT_e}{dt} = -\frac{3\lambda \Omega_0^3}{\hbar \pi k_B^2} \frac{n_e - n_p}{T_e} + \frac{P}{C_e}, \qquad (4)$$

$$\frac{dT_p}{dt} = \frac{C_e}{C_p} \frac{3\lambda\Omega_0^3}{\hbar\pi k_B^2} \frac{n_e - n_p}{T_e} - \frac{T_p - T_l}{\tau_a}, \quad (5)$$

$$\frac{dT_l}{dt} = \frac{C_p}{C_l} \frac{T_p - T_l}{\tau_a}, \quad (6)$$

where $\tau_a$ is the characteristic time for the anharmonic coupling of the hot phonons to the lattice (in this case equals 2.8 ps), $n_e$ and $n_p$ are the electron and hot-phonon distributions given by $n_{e,p} = (e^{\Omega_0/k_B T_{e,p}} - 1)^{-1}$, and $P$ is the laser fluence function; a ratio of $10^3$ between the electronic specific heat $C_e$ and the lattice specific heat ($C_p$ and $C_l$) is known (*30*).

In our calculations, the values of the parameters were chosen to be the same as in ref. (*30*), except for the excitation source which in our case has a fluence of 20 mJ/cm$^2$ and duration of 120 fs. The fit of the simulated lattice temperature to our data (see Fig. 9 A, D, E) gives the following results for the electron-phonon coupling constant in the different samples: $\lambda_{[110]} = 0.12$, $\lambda_{[010]} = 1.0$ and their average $\lambda_{\text{avg}} = 0.56$ in underdoped Bi2212, Fig. 9 D; $\lambda_{[110]} = 0.08$, $\lambda_{[010]} = 0.55$ and their average $\lambda_{\text{avg}} = 0.31$ in optimally doped Bi2212, Fig. 9 A; $\lambda_{[110]} \approx \lambda_{[010]} = 0.40$ in optimally-doped Bi2223, Fig. 9 E. In our procedure, the accuracy in determining $\lambda$ depends on the precision in estimating the decay constant of the Debye-Waler factor, which can be very high given the signal to noise ratio achieved in our experiments. However, the absolute error in the determination of $\lambda$ also depends on the approximations behind the three-temperature model. The best estimate of the likelihood of these numbers comes from the comparison with other techniques and calculations. The average value at optimal doping is in good agreement with the results ($\lambda = 0.26$) of ref. (*30*), which angularly integrates the photoemission among different crystallographic directions. It is also in agreement with "frozen-phonon" calculations (*44*).

The rate of diffraction change provides the time scales of selective electron-phonon coupling and the decay of initial modes involved. The analysis of the first derivative of the Debye-Waller decay helps distinguishing the different processes involved in the decay of the initial excitation, *i.e.* electron-phon coupling and anharmonic phonon-phonon interactions. In Fig. 9 B, the derivatives of the diffraction intensity as a function of time, d$I(t)$/d$t$, are displayed for different polarizations. The presence of a clear inversion point reflects the two processes involved, the one associated with the coupling between excited carriers and optical phonons, and the second that corresponds to the decay of optical modes, by anharmonic coupling into all other modes. The minimum in the derivative, signaling the crossover between these two processes, shifts toward an earlier time when the polarization becomes along the Cu–O bond. In Fig. 9 C, the derivative of the simulated lattice temperature within the three-temperature model, d$T_l(t)$/d$t$, shows a similar two-process behavior. The shift of the minimum to an earlier time can be reproduced by varying the electron-phonon coupling parameter $\lambda$; in contrast, a change in the anharmonic coupling constant $\tau_a$ does not affect the early process, and the corresponding time of the derivative minimum has little shift (Fig. 9 C, *inset*). Thus, consistent with the results of Fig. 9 A, this analysis suggests that the anisotropic behavior of the diffraction intensity is due to a directional electron-phonon coupling.

The derivative minima occur at times of ~1.0, 2.0 and 3.5 ps, respectively, for the polarization at 0°, 22° and 45° with respect to the Cu–O bond direction (Fig. 9 B). The initial rate of the electron-phonon scattering can be obtained through the equation (*42*):

$$\frac{1}{\tau_{el-ph}} = \frac{3\hbar\lambda\langle\omega^2\rangle}{\pi k_B T_e}\left(1 - \frac{\hbar^2\langle\omega^4\rangle}{12\langle\omega^2\rangle(k_B T_e)(k_B T_l)} + \cdots\right) \approx \frac{3\hbar\lambda\langle\omega^2\rangle}{\pi k_B T_e} \quad (7)$$

where $\tau_{el\text{-}ph}$ is the characteristic coupling time constant and $\omega$ is the angular frequency of the coupled modes. Given the values of $\lambda$ (0.55, 0.18 and 0.08 in Fig. 9 C), we obtained $\tau_{el\text{-}ph}$ to be 290 fs, 900 fs and 2.0 ps with an initial $T_e$ = 6000 K and $T_l$ = 50 K. In ref. (*30*), $\tau_{el\text{-}ph}$ was reported to be 110 fs for $T_e$ ~ 600 K. Given the difference in fluence, hence $T_e$, the values of $\tau_{el\text{-}ph}$ obtained here (see Eq. **7**) are in reasonable agreement with the average value obtained in ref. (*30*). It should be emphasized that within such time scale for the electron-phonon coupling, the lattice temperature $T_l$ remains below $T_c$; in Fig. 9 A, the temperature crossover ($T_l > T_c$) occurs at 2 to 3 ps. We also note that at our fluence the photon doping has similar charge distribution to that of chemical doping (*35*).

The influence of polarization on the (00) diffraction rod (which gives the structural dynamics along the *c*-axis) reveals the unique interplay between the in-plane electronic properties and the out-of-plane distortion. Among the high-energy optical phonons that are efficiently coupled at early times, the in-plane breathing and out-of-plane buckling modes are favored (Fig. 5 A) (*7*, *8*) because of their high energy and involvement with carrier excitation at 1.55 eV. Our observation of a faster *c*-axis dynamics when the polarization is along the Cu–O bond implies a selective coupling between the excitation of charge carriers and specific high-momentum phonons. A plausible scheme is the stronger coupling between the antinodal ([010]) charge carriers and the out-of-plane buckling vibration of the oxygen ions in the Cu–O planes. More details will be discussed in the following sections.

**III c. Implications regarding the material`s phase diagram**

Time-resolved electron diffraction provides the opportunity to examine the separate contributions of electronic and lattice-heating to the temperature dependence of electron-phonon coupling. The effect of the equilibrium-temperature can be studied varying the initial sample temperature, and at identical laser fluence, when the electronic temperature rise remains unchanged, the lattice temperature can be tuned between 40 K and higher temperature, thus reaching different points of the phase transition region. If the laser fluence is varied instead, the temperature reached by the out-of-equilibrium electrons can be varied by about one order of magnitude (for fluences between 2 and 20 mJ/cm$^2$), whereas the lattice-temperature change is much slower, and of lower value. The results from such different experiments are given in Fig. 10 A and C.

In Fig. 10 A, where the fluence varied, the overall intensity decay changes significantly, and a faster decay is observed at higher fluences. The overall characteristic time of an exponential fit to the data is displayed as black squares in the inset of Fig. 10 A. As remarked before, these transients are the results of two processes: (i) the ultrafast electron-phonon coupling, and (ii) the slower anharmonic decay of the hot phonon into thermal vibrations; see Fig. 7 B, C. In order to separate these two contributions, we also plot in Fig. 10 B the derivative of the intensity decay for different fluences. The time-constant associated with the electron-phonon coupling is seen to vary modestly as a function of the fluence. The time scale corresponding to the minimum in the derivative is also plotted in the inset of Fig. 10 A. From the comparison between this time-scale and the longer one obtained from the single exponential fit we conclude that the electronic-temperature rise mainly affects the process caused by anharmonic coupling. We also notice that the fluence dependence of both time constants is not monotonic and shows an anomaly in the proximity of the fluence value inducing a lattice temperature rise similar to $T_c$.

In Fig. 10 C, the dependence of the intensity decay rate as a function of the lattice temperature is displayed. In this case, the electronic temperature rise is constant, and so is the anharmonic coupling. In general, a faster decay is observed at higher temperatures, and the trend is understood in view of the two types of phonons present at high temperature: those created through carrier-phonon coupling (low-temperature) and the ones resulting from thermal excitation. This behavior with temperature is consistent with the optical reflection studies made by Gedik *et al.* *(33)*. Also in this case we note that the temperature dependence of the decay rate is not monotonic when the light is polarized along [10]. The anisotropy observed in the Bi2212 samples is apparent at low temperature. Whether or not the anisotropy is related to $T_c$ cannot be addressed with the current temperature resolution, although a correlation is suggested by the data.

The observed temperature dependences suggest that the electron-phonon coupling parameter, to large extent, is insensitive to the electronic temperature, while it could be influenced by the lattice temperature. The effect of heating on the electronic structure is expected to mainly broaden the electronic density of states, which is a determinant of the electron-phonon coupling. However, if there is no strong peak in the density of states at the Fermi level (as is the case for BSCCO cuprates), one may not expect a large effect for the electronic temperature on λ, consistent with our observation.

In Fig. 10 D, the doping dependence of $\lambda$ and the anisotropy observed for different polarizations, $\Delta\lambda = \lambda_{[010]} - \lambda_{[110]}$ (obtained from repeated experiments on different samples and cleavages), are displayed, together with the qualitative trend of the upper critical field (Nernst effect) and coherence length (*45*). The similarity in trend with the upper critical field behavior, which can be related to the pair correlation strength, is suggestive of lattice involvement especially in this distinct phase region where the spin binding is decreasing. In view of an alternative explanation for the doping dependence of the critical field (*46*), our observation of an anisotropic coupling for different light polarizations may also be consistent with the idea of a dichotomy between nodal and antinodal carriers, with the latter forming a charge-density wave competing with superconductivity (*47*). Future experiments will be performed for completing the trends up to the overdoping regime for different superconductor transitions (*48*).

In the 3-layered sample, Bi2223, a much weaker, if not absent, anisotropy was observed at optimal doping (see Fig. 7 D). The electron–phonon coupling in Bi2223 is thus similar for both directions (λ = 0.40) (see Fig. 9 E), signifying that the out-of-plane buckling motions are coupled more isotropically to the initial carrier excitation, likely due to the somewhat modified band structure (e.g., larger plasma frequency; see ref.(*49*)) from that of Bi2212. This observation is consistent with the more isotropic superconducting properties of Bi2223 (*39*). The screening effect for the inner Cu-O layer by the outer ones in Bi2223 (*50, 51*), and the less structural anisotropy between the in-plane and out-of-plane Cu-O distances (*39*), may also play a role in the disappearance of the anisotropic electron–phonon coupling. It is possible that the anisotropy of the excited carriers depends on the number of layers, as will be discussed in the next section.

**III d. Anisotropy with polarized excitations**

In this section we discuss qualitatively different microscopic possibilities that could explain the observed anysotropy. We detail our speculations simulating the polarized optical excitation in a model system (the stripe ground state of LSCO). A rigorous discussion of this issue would require an unambiguous ab-initio description of the electronic properties of doped BSCCO, which to date is still lacking. It is now a well established fact that doped holes in some cuprates self-organize in antiferromagnetic (AF) domain walls (*52-56*). These quasi one-dimensional (1D) structures called stripes where predicted by mean-field theories (*57*) inspired by the problem of solitons in conducting polymers (*58*). In some compounds, stripes are clearly observed and are accompanied by a spontaneous braking of translational and rotational symmetry in the Cu-O planes. For example, in 1995 Tranquada and collaborators observed a splitting of both spin and charge order peaks in $La_{1.48}Nd_{0.4}Sr_{0.12}CuO_4$ by elastic neutron scattering (*52*). The outcome of this experiment resembled the observation made in the nickelates, where both incommensurate antiferromagnetic (AF) order (*59, 60*) and the ordering of charges have been detected by neutron scattering and electron diffraction, respectively (*60, 61*). In Ref. (*60*), it was shown that the magnetic ordering in $La_2NiO_{4.125}$ manifests itself as occurring first and third harmonic Bragg peaks, whereas the charge ordering is associated with second harmonic peaks. From this, it was concluded that the doped holes arrange themselves in quasi-onedimensional structures, which simultaneously constitute antiphase domain walls for the AF order. There are several compounds, however, where static long-range order has not been observed leading to the speculation that stripes survive as a dynamical fluctuation. An interesting possibility is that there exist precursor phases of the stripe phase where translational symmetry is preserved but rotational symmetry is spontaneously broken; this, in analogy with liquid crystals, has been called nematic order (*62*). Stripe phases, breaking fundamental symmetries of the lattice, lead to the appearance of new collective modes that do not exist in a normal Fermi liquid. These collective modes may play an important role in the mechanism of superconductivity as pairing bosons. In our experiment we do not detect a spontaneous symmetry breaking however our results are compatible with the proximity to a nematic phase as explained below.

Although the three-temperature model gives a reasonable description of the relaxation times for each polarization, the basic assumptions of the model have to be re-examined when considering the anisotropy itself. In the standard formulation, one assumes that the electronic relaxation time (< 100fs) is much shorter than the electron-phonon coupling time (few hundreds fs). The fact that one observes an anisotropy on the ps time scale suggests that this assumption brakes down, because electrons reaching thermal equilibrium in ~ 100 fs would not have a memory of the excitation direction at later times. This means that some anisotropic electronic state is excited by the laser, which has a longer relaxation time in the nodal direction when compared with that of the antinodal direction. Anisotropic scattering rates are well documented in photoemission spectroscopy of cuprates (*63*). The anomalous long electronic relaxation time suggested that the exited state is a low lying collective electronic excitation, where the decay rate is limited by Fermi statistics and many-body effects, rather than by an interband or other high-energy excitation. Within the Fermi liquid theory, low energy excitations can be characterized by the oscillatory modes of the Fermi surface.

The lowest energy anisotropic excitations are the Pomeranchuk modes, which we illustrate schematically in Fig. 11 C, D for a cuprate Fermi surface. The thick red line is the undisturbed Fermi surface and the thin blue line is a snapshot of the oscillating Fermi surface. Thus, depending on the laser excitation direction the Fermi surface can remain oscillating in the antinodal direction (C) or the nodal direction (D). If the system is close to a Pomeranchuk instability, the relaxation times of these two nonequilibrium configurations can be very different (*64, 65*). A Pomeranchuk instability along the nodal direction would make that particular relaxation time very long and is one possible explanation for our results. After crossing the instability point the deformation becomes static and the system acquires nematic order, i.e. brakes the $C_4$ symmetry of the lattice along the diagonals without braking translational symmetry. This, however, is in contrast with the tendency of cuprates to brake $C_4$ symmetry along the Cu-O bond (except for slightly doped LSCO) through charge ordered states, termed stripes. In Fig. 13 B we display static stripe order running along the y direction, according to a microscopic computation of stripes in the three band Hubbard model (*66*). Stripes are domain walls of the antiferromagnetic order where holes tend to accumulate. In cuprates, static charge order as depicted is only seen under very special conditions which favor stripe pining, more often stripes are believed to be dynamical objects. Indeed, the excitation spectrum of cuprates above some minimum energy $\omega_0$ coincides with the excitation spectrum predicted by the stripe model even if static stripes are not detected (*67*). The given $\omega_0$ can be interpreted as the energy scale above which stripes look effectively static. Stripes are good candidates to disrupt the Fermi liquid ground state and be responsible for the peculiar properties of cuprates. Based on photoemission experiments (*63, 68, 69*), it has been proposed (*70*) that the Fermi surface has a dual nature: fluctuating stripes produce a blurred "holy cross" Fermi surface, see Fig. 11 B, while low energy quasiparticles average out the stripe fluctuations and hence propagate with long relaxation times along the nodes. In Fig. 11 B, the two resulting structures are schematically shown, the Fermi surface is obtained from a calculation of static stripes on LSCO and has been artificially blurred to simulate the fluctuating character. This blurred FS coexists with the sharp FS due to nodal quasi particles, indicated by the thin lines.

Another possible explanation of the anisotropy found for the scattering rates is that the Pomeranchuk Fermi surface modes have intrinsic long relaxation times, but the relaxation time along the antinodal direction becomes shorter because of scattering with stripe fluctuations which are known to be along the Cu-O bond. In this scenario, when the electric field is on the diagonals, the slowly relaxing nodal states are excited, but when the electric field is parallel to the Cu-O bond internal excitations of the stripe are produced and relax fast due to electron-electron scattering and strong coupling to the lattice. This is consistent with the observation by photoemission that antinodal carriers (charges along the Cu-O bond) are strongly coupled to out-of plane phonon modes of the oxygen ions. In fact, in diffraction, this would result in a faster decay of the Debye-Waller factor when more charges are involved along the Cu-O direction.

We now discuss the mechanism by which the Pomeranchuk modes are excited. This mechanism must necessarily involve more than one photon since single photon absorption is described by linear response theory where the response along the diagonals can be simply decomposed in the sum of the responses along the bonds. The possible path that would excite the Pomeranchuk modes is depicted in Fig. 11 A, a sort of Raman or two step process. First a photon is absorbed by a dipole allowed transition. At a latter time another photon is emitted leaving the system in one of the two possible excited states with different lifetimes. This scenario would require the presence of resonant absorption states around the laser excitation energy (1.5 eV), as the triplet structure observed in the pair-breaking spectroscopy on YBCO sample for example (*71*), however a thorough assignement of the absorption features in different cuprates optical spectra is difficult and still underway, as we discuss below. Dynamic stripes may be involved in this process too since their anisotropic character make them couple well to the Pomeranchuk modes. For this to be possible there should be a stripe absorption mode at the energy of the incoming photon. In order to substantiate this view, we present in what follows computations of the optical absorption of the stripes relevant for the first step of the process and compare with experimental results.

The in-plane optical absorption of different cuprates superconductors (Bi2212, Bi2223 and LSCO) is displayed in Fig. 12. In panel C, the spectrum at different temperatures for Bi2212 is displayed. A large metallic component is found at low frequency, often referred to as the Drude peak (*15*). Most of the temperature dependence is observed in this part of the spectrum, and in the inset of Fig. 12 B, one can see the effect of the opening of the superconducting gap below 100 meV in the optical spectrum.

At higher energy, above the plasma edge of the material (> 1eV), several absorption features are observed. In Fig. 12 B, the spectrum is decomposed into different components by a standard Drude-Lorentz fit, and one can see that a feature centered around 1.8-2 eV is obtained (evidenced by a thick blue trace). This absorption is often ascribed to a charge transfer excitation between Cu and O ions in the ab-plane of the material, but in BSCCO, its assignment is complicated by the presence of other strong interband transition in that spectral region. However, since the electronic states close to the Fermi energy in all cuprates are Cu $3d_x^2{}_{-y}^2$ orbitals hybridized with O $2p$ orbitals, most of the temperature dependence is expected to be found in spectroscopic features associated with these orbitals.

In Fig. 12 A, the difference between spectra at different temperatures (T = 280 K – T = 20 K) is noted to be large in the Drude region of the metallic carriers (inset of Fig. 12 A), and peaks in correspondence to the charge-transfer peak around 2 eV. These features are quite common to cuprates. The reason is that the low energy electronic structure of these materials is dominated by the physics of the Cu-O plaquettes, which are common features of high-temperature superconductors. To emphasize this point, we display, in Fig. 12 C, the spectra of different chemical compositions, Bi2212, Bi2223, and LaSrCuO. The overall shape of the spectrum is similar in all cases, and the first absorption feature above the plasma edge is always attributed to Cu-O charge transfer excitations.

LSCO is an ideal material for the theoretical investigation of the optical spectrum, because it has a single Cu-O layer per unit cell and its overall crystal structure is among the simplest of all cuprates. Also, experimental evidence of stripes has been reported in this material (*18*). In Fig. 12 C, we show the comparison between the theoretical optical absorption spectrum of LSCO based on the metallic stripes model and the experimental one. The main features of the spectrum are reproduced by theory, and the 1.2-1.5 eV absorption feature is assigned to a charge transfer mode inside the stripe, which again involves the motion of a charge along the Cu-O bond. This is expected to be the mode responsible for the first step of the Raman process. We investigated LSCO because of the good success of these calculations in reproducing the optical spectrum of the material, because of its inherently simpler structure. We expect the situation in BSCCO to be not very different, again because we focus on features mainly related to the Cu-O plane.

In BSCCO and LSCO cuprates, the in-plane linear optical absorption of light can be considered as nearly isotropic (*15, 72*). Our observation of an anisotropic structural dynamics as a result of light excitation polarized in different directions within the plane suggests that the out-of-equilibrium electronic structure is capable, during its thermalization, to induce distinct structural changes which distort the lattice in a way that is observable at longer times. According to our computations (*67*), when the electric field is parallel to the Cu-O bond the 1.5 eV polarized light couples strongly with the stripe excitation. In Fig. 13 A and B, the charge variation induced by 1.2 eV light polarized parallel (panel A) and perpendicular (panel B) to the charge stripe is shown. The red arrows indicate charge increase (up arrow) or charge decrease (down arrow) in a certain area, whereas the green arrows indicate the current induced by light absorption. When light is polarized perpendicular to the stripes (which would correspond to one of the two directions of Cu-O bonds, (10) for example), a charge transfer mode within the stripe is excited. This is visible as red arrows in Fig. 13 C, and indicate a charge increase on the oxygen site and a charge decrease on the neighbouring copper site.

When light is polarized parallel to the charge stripe (along the other Cu-O direction, (01) for example), no such charge transfer is observed. For light polarized at 45 degrees, we expect that the nodal quasiparticles are excited through other intermediate states (note that they are not considered in the present calculations for technical reasons) and long relaxations times are found. In Bi2223, where the anisotropy of the electron-phonon coupling was found to be much weaker, a weaker stripe charge ordering may be present due to the different doping of the layers and the increased three dimensional properties of the system. It would be interesting to study in more detail the doping and temperature dependence of the anisotropy to verify if there is a precise point in the temperature-doping plane where a Pomeranchuk instability manifests as a divergence of the relaxation time, or if the phenomenon is more related to stripe physics. Another important check would be to study the dependence of the result on the laser excitation energy to verify the importance of the 1.2-1.5 eV stripe absorption band.

**III e. Nonequilibrium phase transitions in LCO: the structural isosbestic point**

As discussed above, charge transfer excitations can play a crucial role in the dynamics of cuprates. In $La_2CuO_4$, manifestations of structural dynamics can be observed when the electrostatic imbalance induced by charge transfer excitations causes lattice changes in the direction perpendicular to the Cu-O planes.

The dynamical behavior of LCO structure is displayed in Fig. 14. Here, shown are the time resolved diffraction difference images at room temperature, displaying the changes induced by the excitation pulse with a fluence of 20.6 mJ/cm$^2$. These frames were obtained at the specified times and referenced, by subtraction, to a frame at negative time. White regions indicate intensity increase, while dark regions relate to intensity decrease. After the excitation, Bragg spots move down vertically, as evidenced by the appearance of white spots below dark regions in the difference frames. No substantial movement is observed along the horizontal direction. The changes are maximized around ~ 120 ps and relax on a longer time scale (~ 1 ns).

In Fig. 15, the profiles of the 0010 and 008 Bragg spots along the *c*-axis direction are shown at different times after the laser excitation (in this probing geometry, 006 is not shown since, it shifts below the shadow edge at 112 ps and partially disappears for geometrical reasons). Before the laser excitation, both Bragg spots are centered at the equilibrium values, but at 112 ps after the excitation, they are centered at smaller s values (upper curves in Fig. 15 A show $\Delta s/s = -2.5\%$ for a fluence of 20.6 mJ/cm$^2$). In between these two time frames, the evolution of the Bragg spot profiles is not a continuous shift of its center position.

Rather, all the curves obtained at different time delays cross at a certain *s*-value. Furthermore, the total intensity underneath each Bragg spot stays constant within 2-3% during the entire timescale of the experiment. This behavior is very different from all other studied materials, and from that of BSCCO as well. In GaAs (*73*), for example, the center position of the Bragg peaks shifts continuously to a lower momentum transfer value, indicating a continuous expansion along the surface normal direction, and the total intensity underneath the profiles decreases because of the Debye-Waller effect. In BSCCO samples (*24*), the very large value of the *c*-axis, and the presence of a heavy element like bismuth, causes the diffraction to be of a rod-like shape, preventing careful analysis of the position changes. However, the main effect of light excitation was found to be the Debye-Waller decrease of the diffraction intensity. In LCO, electrons probe the specimen in transmission at a lower scattering vector with respect to the experiments on BSCCO, giving a smaller Debye-Waller effect, according to Eq. (1). Also, the temperature jump induced by pump pulses in LCO is significantely smaller than in BSCCO, again limiting the magnitude of the average atomic displacement. These facts, together with the nanometric size of the probed domains, could play a role in reducing the Debye-Waller effect below our observation capabilities.

The conservation of the total intensity and the existence of a crossing point (as clearly seen in Figs. 15 A, and 16 A) is not consistent with a continuous expansion; it rather indicates a direct population transfer between two phases of the lattice with different *c*-axis constants. Such a crossing behavior in optical absorption spectroscopy would be termed "isosbestic point", corresponding to the spectral position where the two interconverting species have equal absorbance; regardless of the populations of the two states, the total absorption at the isosbestic point does not change if the total concentration is fixed. In the present case, we term this point as "structural isosbestic point", corresponding to the point in the momentum space where the two structures are contributing equally to the diffraction intensity.

The relaxation process back to the equilibrium value of this new phase follows a different dynamic. The lower trace of Fig. 15 A shows the evolution of the Bragg spot profiles between 112 ps and 1217 ps. In this period, no crossing point is observed and the center of the Bragg spot shifts continuously back to the equilibrium value. The timescale of the return to equilibrium is also slower by about an order of magnitude (about 300 ps as opposed to 30 ps).

Structural distortions give changes in diffraction obeying the scaling relation in Eq. 2. We test this scaling relation, as well as the presence of spurious motions of the electron beam, in order to verify that the observed dynamics originate from atomic motions.

In Fig 15 B, we compare the relative changes in the position of 008 and 0010 Bragg spots. We also plot the change in the position of the undiffracted direct beam. The center position of each spot is obtained by fitting the vertical profile into a Gaussian form. Between 0-112 ps, the Bragg spots can not be described by a single Gaussian curve since more than one phase with distinct structural parameters coexist. In this time period (shown by a transparent yellow strip), a fit to a single Gaussian can not adequately describe the profiles. Outside this region, a single Gaussian can fit the data properly.

Since the direct beam position does not change with time, we excluded the possibility that the observed behavior come from the shift of the entire pattern or that surface charging is contributing to the diffraction. Moreover, for the structure, the change in different orders should scale according to the order number, i.e. $\Delta s/s = -\Delta c/c$, where $\Delta s$ is the shift of the n-th order Bragg spot (as was verified for BSCCO as well). This means that the 008 Bragg spot should move only by 80% of the 0010 Bragg spot. The dashed blue curve in Fig 15 B shows the shift of the 0010 Bragg spot scaled by 80%. Furthermore, the agreement between this and the shift of 008 spot (red curve) confirms that the observed dynamics are due to real structural changes.

The dependence on the excitation density of the observed phase transition reveals detail of the interplay between the lattice and the electronic structure. In Fig. 16 A, the generation of the new phase between 0 and 112 ps is depicted, whereas in Fig. 16 B we show the relaxation back to the ground state for different laser fluences at room temperature. First, we see that at 2.8 mJ/cm$^2$ there is no observable change. Above this intensity, there is a crossing point observed in Fig. 16 A, whereas the peak position shift continuously back to the equilibrium value in Fig. 16 B. The maximum change in the position of the Bragg spot increases with increasing fluence (Fig. 16 A). The characteristic time of this process does not depend on the laser fluence and is ~30 ps.

In Fig. 17 B, the maximum c-axis expansion, $\Delta c$, obtained on the time-scale indicated by the dotted line in Fig. 17 A, is displayed as a function of laser fluence. These data are reported for two temperatures, 20 and 300 K, respectively. For these measurements, we have used a polarizer and a half-wave plate in order to be able to adjust the laser fluence in fine steps. This arrangement enabled us to change the laser fluence continuously without changing the spatial overlap or the relative arrival times of the laser and electron pulses. At each laser fluence, we obtain $\Delta c$ by recording two profiles at times −85 ps and 130 ps. Below a threshold intensity of ~5 mJ/cm$^2$, no change was observed. Above this threshold intensity, $\Delta c$ grows linearly with increasing fluence.

In order to better understand the microscopic meaning of this threshold fluence, we consider the number of photons absorbed per copper site for each fluence. The energy ($u$) deposited into the cuprate film per unit volume for each pulse (in the surface region probed by the electron beam) is given by $u = F_0(1-R)\alpha$, where $F_0$ is the incident laser fluence, $R$ is the reflectivity of the cuprate film, and $\alpha$ is the absorption coefficient of the film. The energy absorbed per unit cell ($u_c$) is given by $u_c = u \cdot v_c$, where $v_c$ is the volume of the unit cell. The number of photons absorbed per copper site ($\delta_p$) can be calculated by dividing the energy absorbed per unit cell by the energy of each photon ($\hbar\omega = 1.5\ eV$), after taking into account that there are two copper atoms per unit cell, i.e. $\delta_p = F_0 v_c (1-R)\alpha /(2\hbar\omega)$. Given the values of $R = 0.1$, $\alpha = 7 \times 10^4$ cm$^{-1}$ ($R$ and $\alpha$ were obtained from Figs. 12 and 9 of ref. (*74*), respectively, measured for a similar sample) and $v_c = a\ x\ b\ x\ c = 186.12$ Å$^3$, we obtain $\delta_p = F_0 \times (24.4\ cm^2/J)$. Using this expression, the laser fluence can be converted into the number of photons absorbed per copper site, as shown in the top horizontal axis of Fig. 17 B. It is intriguing that the threshold intensity corresponds to ~0.1 photons absorbed per copper site. This is very close to the number of chemically doped carriers needed to induce superconductivity.

As far as the temperature dependance is concerned, only a slight increase in the rise time can be seen as the sample is cooled down, similar to what was observed also in BSCCO.

**III f. Theoretical modeling of the light-doped phase transition**

The structural dynamics observed in the cuprate film show several distinct features which need to be addressed by a theoretical model. These are a large increase of the *c*-axis constant, the existence of a structural isosbestic point at intermediate times, a continuous shifting of the Bragg spot profiles in the relaxation regime, and the existence of a threshold fluence and a linear dependence of the expansion on the laser fluence above this threshold. Furthermore, the characteristic timescales involved (30 ps for the onset and 300 ps for the relaxation) do not strongly depend on temperature and fluence.

Below, we will present a simple energy landscape model that can account for these findings.

**Thermal expansion model**

A simple thermal expansion mechanism can not explain the experimental observations for three reasons. First, a 2.5 % increase in the *c*-axis lattice constant would correspond to an unphysical 2500 K rise in the lattice temperature, since the linear thermal expansion is $\alpha_l \leq 1.0 \times 10^{-5}$ K$^{-1}$ (*75*). Second, in the thermal expansion scenario, the total intensity underneath a Bragg spot is expected to decrease due to the Debye-Waller effect, caused by the phonon generation following the photoexcitation, whereas in our case the total integrated intensity underneath a Bragg spot is found to be nearly constant. Finally, the thermal expansion model would predict a monotonic shift of the Bragg spots into lower momentum transfer values, whereas we observe a crossing point that can not be explained by a continuous increase of the interplanar distance.

**Structural changes and the in-plane charge transfer**

It is worth considering some absorption characteristics. The substrate (LaAlSrO$_4$) does not absorb at our laser wavelength of 800 nm. The penetration depth of the laser beam in the cuprate film was estimated to be 143 nm, given the absorption coefficient of $\alpha = 7 \times 10^4$ cm$^{-1}$ (*74*). The electron beam is at a grazing incidence angle of 1.5$^0$, and due to the small mean free path of the high energy electrons, and the low incidence angle, only the top few nm of the 52 nm thick film can be probed. Therefore, only the cuprate film can contribute to the dynamics and the substrate is not expected to have any direct role.

As shown in a previous section, a charge transfer excitation in cuprates is expected around 2 eV. This excitation involves a charge transfer from the 2*p* orbital of oxygen into the 3*d* orbital of copper (*15*). Our excitation energy (1.55 eV) falls in the proximity of such charge transfer. It should be noted, however, that this assumption is model dependent since a consensus has not been reached yet on the assignment of the different absorption features in cuprates.

LCO is a highly ionic compound with a large cohesive energy. As a result of a charge transfer excitation, and subsequent changes in the valency of the in plane copper and oxygen, a weakening of the coulomb attraction between the planes is induced, which leads to expansion. In the experiment, we observe that almost the entire Bragg spot, and not a fraction of it, undergoes the change, indicating that macroscopic scale domains (which define the coherence length of the Bragg diffraction) are involved in this phase transformation. Above a certain fluence, charge transfer excitations are shared among multiple unit cells and macroscopically-sized domains are created with distinct electronic and structural properties.

**Cohesion energy calculations**

In order to quantitatively describe the effect of the in-plane charge transfer on the lattice structure, we calculated the cohesion energy as a function of structural parameters both in the ground and in the charge transfer state; we shall refer to the latter as the excited state. In order to model the excited state at each fluence, we calculated the number of photons absorbed per copper atom ($\delta_p$) as displayed on the top scale of Fig. 17 B by using the absorption coefficient and the carefully measured laser fluence. When the charge transfer excitations are shared uniformly across the Cu-O planes, the valance of in-plane Cu atom changes from +2 to +2 - $\delta_p$ and the valance of oxygen changes from -2 to -2 + $\delta_p/2$. We then calculate the cohesion energy as a function of the structural parameters for each value of $\delta_p$, and find the values of these parameters that minimize its energy (Fig. 18). Below we will describe the results of two independent calculations.

In the first of these calculations, we used the model of Piveteau and Noguera (*76*) to compute the cohesion energy. It was originally used to reproduce the structural parameters of LCO at equilibrium. The cohesion energy is expressed as the sum of pair-wise interaction between the atoms, taking into account three microscopic terms: the direct Coulomb interaction treated in the point charge approximation, a hard-core repulsion of the Born-Mayer type, accounting for the orthogonality of the atomic orbitals on different atoms at short distances, and the van der Waals term. The interaction energy ($E_{ij}$), of two atoms i and j, having charges $Q_i$ and $Q_j$ at a distance $R_{ij}$ is given by

$$E_{ij} = Q_i Q_j / R_{ij} + B_{ij} \exp(-R_{ij}/\rho_{ij}) - C_{ij}/R_{ij}^6 \quad \textbf{(8)}$$

This expression contains three sets of parameters ($B_{ij}$, $C_{ij}$, $\rho_{ij}$) which are obtained from the atomic values ($B_{ii}$, $C_{ii}$, $\rho_{ii}$) using the following empirical expressions: $B_{ij} = (B_{ii}B_{jj})^{1/2}$, $C_{ij} = (C_{ii}C_{jj})^{1/2}$, $2/\rho_{ij} = 1/\rho_{ii} + 1/\rho_{jj}$. These expressions are based on the fact that the van der Waals interactions are related to the product of the atomic polarizabilities and the hard-core repulsion involves a product of exponentially decreasing atomic wave functions. In our calculation, we used the same parameters that were invoked in order to reproduce the equilibrium structure (*76*). In order to obtain these parameters, Piveteau and Noguera used the nine atomic values ($B_{ii}$, $C_{ii}$, $\rho_{ii}$) for $Cu^{2+}$, $O^{2-}$, and $La^{3+}$ from the literature (*77*), and adjusted them in such a way that they reproduced the more simple structures of CuO and $La_2O_3$.

The total internal energy of the crystal ($E_T$) was obtained by summing the internal energy ($E_p$) of the units (defined below) of La$_2$CuO$_4$, and the interaction energy $E_{PQ}$ between two units P and Q as

$$E_P = \frac{1}{2}\sum_{i,j} E_{ij} \ (i \in P, j \in P, i \neq j) \quad (9)$$

$$E_{PQ} = \sum_{i,j} E_{ij} \ (i \in P, j \in Q) \quad (10)$$

The total internal energy of N units is given by

$$E_T = \sum_P E_P + \frac{1}{2}\sum_{P,Q} E_{PQ} \ (P \neq Q) \quad (11)$$

where the sums go over N units. The energy per unit ($E$) is given by

$$E = E_T/N = E_{P_0} + \frac{1}{2}\sum_{Q(\neq P_0)} E_{P_0 Q} \quad (12)$$

where $P_0$ denotes any central unit.

For rapid convergence, we grouped the atoms into elementary units without dipolar moments (*76*), as shown in Fig. 18 A. It includes a CuO$_6$ octahedron and two La atoms. The oxygen atoms in the CuO$_2$ planes are labeled as O(2) and out of plane oxygen atoms are labeled as O(1). The O(2) atoms are counted as ½ in the summations, since they are shared by two units. La$_2$CuO$_4$ in the tetragonal structure can be described by four structural parameters: the distance between CuO$_2$ planes (c/2), Cu – O(1) distance ($d_1$), the Cu – O(2) distance ($d_2$), and the Cu – La distance ($d_3$). We calculated the internal energy per elementary unit as a function of these four parameters $E(c, d_1, d_2, d_3)$ by summing up ~40,000 interactions. The minimum value was found to be *E (c = 13.1315 Å, $d_1$ = 2.4004 Å, $d_2$ = 1.897 Å, $d_3$ = 4.7694 Å) = -173.8462 eV*. These values are in very good agreement with the results of Piveteau and Noguera $E$ $(c = 13.15$ Å, $d_1 = 2.399$ Å, $d_2 = 1.897$ Å, $d_3 = 4.7780$ Å$) = -173.83$ $eV$.

The effect of inplane charge transfer was accounted for in the following way. We assumed that the excitations are shared uniformly to create domains of modified valancy of Cu and O(2) atoms in the $CuO_2$ planes. If the number of photons absorbed per Cu atom is $\delta_p$, then the valancy of Cu atoms are changed from +2 to $+2 - \delta_p$, and the valancy of O(2) atoms are changed from −2 to $-2 + \delta_p/2$. The valancy of O(1) and La atoms remain unchanged. Because the cuprate film is epitaxially grown on the substrate, there can not be any change of inplane structural parameters, since they are anchored by the underlying substrate (as confirmed experimentally). Therefore, we assume that the value of the in plane Cu-O(1) distance $d_2$ can not change during the time-resolved measurements and is fixed at its equilibrium value of 1.897 Å.

For a given $\delta_p$, we search for the parameters ($c$, $d_1$, $d_3$) that minimize the energy per elementary unit $E(c, d_1, d_2 = 1.897$ Å, $d_3, \delta_p)$ using two different methods. In the first method, we assume that only a uniform stretching of the unit cell can take place in the $c$-axis direction. We search for the minimum energy configuration by varying $c$, $d_1$ and $d_3$ proportionally. In Fig. 18 B, we present the calculated cohesion energy as a function of the $c$ axis constant for the ground state ($\delta_p = 0$), and excited state with $\delta_p = 0.3$. The absolute value of the cohesion energy in the excited state is correct up to an overall constant because of the uncertainties in the electronegativity and ionization energy of copper and oxygen, respectively. The dashed red curve in Fig. 18 C shows the calculated $\Delta c$ as a function of $\delta_p$. In the second method, we calculate $\Delta c$ (dashed black curve in Fig. 18 C) by allowing the structural parameters ($c$, $d_1$, $d_3$) to change independently.

Besides the calculation mentioned above, independent calculations were performed by our collaborators (*78-79*), which ignore the relatively small van der Waals contribution and uses a somewhat different parameters ($B_{ii}$, $C_{ii}$). The cohesion energy was written as the sum of the Madelung energy and the core repulsion energy. The standard Born-Meier form of $B_{ij} \cdot \exp(-R_{ij}/\rho_{ij})$ was used in order to model the core repulsion energy where the indexes $i$ and $j$ enumerates the relevant nearest-neighbour pairs (O-O, Cu-O, and La-O). The constants ($B_{ij}$, $\rho_{ij}$) in the ground state ($\delta_p = 0$) were first optimized by matching the experimentally determined equilibrium distances and the known lattice elastic constants with the minimum of the cohesion energy and the values of its derivatives. It was further assumed that, since the film is epitaxially constrained to the substrate in the horizontal direction, the in plane lattice constants are fixed by the substrate and could not change, as observed in the experiment. Since there was no stress on the film surface, it can freely expand along the vertical direction (along the *c*-axis). For a given fluence (equivalent to $\delta_p$), the cohesive energy $E(r_1, ..., r_N, \delta_p)$ was calculated as a function of the structural parameters and the new crystal configuration is determined by minimizing the potential energy. We find that the new minimum of $E(r_1, ..., r_N, \delta_p)$ occurs at a higher *c*-axis constant and $\Delta c$ depends linearly on $\delta_p$ as shown by the green line in Fig. 18 C. The agreement between the slope of the green theoretical curve and the slope of experimental data (blue curve) is excellent.

Considering that all the parameters required for these calculations were obtained from equilibrium constants and there were no adjustable parameters to model the time-resolved changes, cohesion energy calculations reproduce the experimentally observed trend remarkably well (Fig. 18 C). The inclusion of van der Waals term in the cohesion energy calculation and/or the usage of slightly different atomic constants ($B_{ij}$, $C_{ij}$, $\rho_{ij}$) do not change the results significantly (about 20% change in the slope of $\Delta c$ vs $\delta_p$).

**The energy landscape**

Based on the above discussion, we can construct a descriptive energy landscape (see Fig. 19). If we consider that multiple coordinates are involved, the observed time-resolved dynamic is the result of the trajectory of motion on this multidimensional energy landscape. In the simplest form, there are two main coordinates that we need to consider: the charge redistribution process in the $CuO_2$ planes, and the macroscopic structural change, mainly expansion along the *c*-axis.

After photoexcitation, the microscopic excitations are formed with no apparent delay. This represents the first step occurring on the ultrafast time scale, and is shown in Fig. 19 as the movement from the initial Franck-Condon region to the modified charge transfer state on the energy surface. This time scale is consistent with the direct movement along a potential energy surface. It is very likely that the initial fast relaxation observed in the time-resolved optical experiments in cuprates is in fact probing this initial step. The timescale for the generation of the macroscopic phase with a longer *c*-axis constant is ~30 ps. This time is too long to be a direct motion along a repulsive potential energy surface. Instead, it involves a transition of a barrier type crossing from the initial local minimum to the macroscopic minimum stabilized by the lattice relaxation. A well defined crossing point also supports this argument. The existence of a structural isosbestic point shows that there is a direct population transfer between the two states which have two well defined values of the *c*-axis constant. The values of the *c*-axis constants in between these two are not observed. This situation can happen if the system is crossing a hill in the potential energy surface during the motion between two local minima. The actual barrier crossing should be very fast, and most of the times it should be found in one of the valleys. The timescale is related to the height of the barrier. A sketch of the excited energy surface showing these initial motions is displayed in Fig. 19.

In the relaxation regime, the *c*-axis constant decreases continuously with a very slow time constant of about 300 ps. This is too slow to be just a movement along an energy surface. It also can not be explained by a single barrier crossing since we observe transition through all the *c*-axis constant values continuously. We should consider the fact that more than one local minimum state may exist as a result of a non-adiabatic "covalent-ionic" interactions between the excited states and ground state energy surfaces. As a result, this might produce a rough energy landscape consisting of small hills and valleys. The observed dynamics might reflect the motion of the system on this kind of an energy landscape.

Another way to understand the slow timescale is to consider the fact that electronic and structural relaxations are coupled. In order for the charges to fully recombine, the lattice has to relax as well, which naturally takes a long time especially if the acoustic phonons are involved. The agreement between the observed 300 ps decay constant in our experiment and the decay time of long lived acoustic phonons seen by the time resolved reflectivity experiments (*80*) also supports the involvement of acoustic degrees of freedom in the relaxation regime. We do not observe strong temperature dependence in this rate between room temperature and 20 K which is also consistent with a scenario dominated by acoustic phonons, as the Debye temperature for this material is 163 K (*81*). The fact that the total intensity of the [0 0 l0] Bragg spots does not change appreciably suggest that the modes involved are mostly in-plane phonons with vibrations orthogonal to the c axis direction.

## IV CONCLUSIONS

The observation of strong charge-lattice interaction in cuprates superconductors, in particular the interplay between the electronic excitations and the *c*-axis motion of the ions, suggest considerations beyond the standard 2D models (*82*) (*21*). Recent theoretical work has incorporated lattice phonons in the *t-J* model to account for the observed optical conductivity (*83*), and new analysis of the optical conductivity on a variety of samples confirmed the scenario of a strong coupling between a bosonic spectrum, consistent with a combination of phonon modes and magnetic excitations (*84*). Moreover, new band structure calculations suggested that large and directional electron-phonon coupling can favor spin ordering (*22*). The anisotropic coupling observed in BSCCO and the intermediate to high values of $\lambda$ obtained imply that the time scale for the scattering of electrons by certain phonon modes can be very fast, comparable to that of spin exchange (40 fs in the undoped phase). As a result, both the magnetic interactions and lattice structural changes should be taken into account in the microscopic description of the pair formation.

In order to explain the microscopic origin of the anisotropy itself, we invoked the charge stripes model. In this model, the material has a non Fermi liquid ground state consisting of ordered charges and spins along particular directions. It is shown that excitations of this ground state can be anisotropic with respect to light polarization, and that a particular charge transfer can be excited within the stripe domain when light is polarized along the Cu-O bond. The interplay between stripes and more conventional quasi-particles could be responsible for the observed anisotropy, as already suggested by photoemission data and theoretical considerations.

The nature of the observed anisotropy, and the symmetry of the carriers involved, also stresses the importance of the interplay between the in-plane charge redistributions and the out-of-plane distortion of the unit cell. Another striking manifestation of this is the structural isosbestic point observed in LCO thin films. In this case, the charge-transfer nature of the optical excitation is used in order to explain, with success, a structural instability along the *c*-axis. Above a certain threshold, a direct conversion between two distinct structural phases of the lattice having different *c*-axis values was formed. The observed structural changes can be explained by a model based on changes in the valances of the in-plane oxygen and the copper which are caused by the charge transfer excitation. The amount of expansion in the *c*-axis constant can be correctly calculated by minimizing the cohesion energy as a function of structural parameters. The value of the threshold fluence corresponds to ~0.1 photons per copper site, which is very close to the number of chemically doped holes required to induce superconductivity. Using the observed time constants and the sequence of distinct structural changes, we presented a picture of the energy landscape and the trajectory that is taken by the system.

**Acknowledgement.** This work was supported by the National Science Foundation and the Air Force Office of Scientific Research in the Gordon and Betty Moore center for physical biology at Caltech. The authors acknowledge stimulating discussions with prof. Z.X. Shen, prof. A. Georges, Dr. Th. Jarlborg, Dr. A. Kuzmenko, Dr. E. Giannini, and prof. D. van der Marel. Dr F. Carbone acknowledges support from the FNS Suisse through an Ambizione grant.

**Figure Legends:**

**Fig. 1.** Experimental set-up. A train of pulses from an amplified Ti:Saph laser is used in order to excite a specimen with polarized light, and to generate electrons from a photocathode. The delay between pump photons and probe electrons is controlled by a motorized delay line. The scattered electrons are recorded by a CCD camera (details in the text).

**Fig. 2.** OpD Bi2212 static diffraction patterns. (A–C): reflection patterns obtained at three different electron probing directions $\vec{v}_e$ (by rotating the crystalline sample), as indicated in the lower right corner. The large lattice constant along $c$ and the nm-depth of electron probing give rise to the rod-like patterns; from (A), the intensity modulation along the diffraction rods gives the out-of-plane lattice parameter of $c = 30$ Å. The indices for different diffraction rods are given. Note that the satellites of the main diffraction rods in (C) manifest the 27-Å modulation along the $b$-axis of Bi2212. (D): Transmission diffraction pattern obtained by our electron microscope. The square in-plane structure is evident, with the presence of the $b$-axis modulation which is also seen in (C). The colored arrow indicates the direction of the cuts taken in the reciprocal space for the analysis of the data.

**Fig. 3.** OpD Bi2223 Bragg peaks. Cuts along the directions indicated by the coloured arrows in Fig. 2 are displayed. (A): cut along the (n00) direction. (B): cut along the (nn0) direction. (C): cut along the (20n) direction. (D): cut along the (0n0) direction. All Bragg peaks are indexed.

**Fig. 4.** OpD Bi2223 and UD Bi2212 diffraction patterns and Bragg peaks. (A): diffraction pattern of optimally doped Bi2223. The electron beam is parallel to the (110) direction. (B): diffraction pattern of under-doped Bi2212. The electron beam is parallel to the (110) direction. (C): cut along the (nn0) direction of Bi2223. (D): cut along the (nn0) direction of under-doped Bi2212. All Bragg peaks are indexed.

**Fig. 5.** Bi2212 unit cell and Bi2223 modulated supercell. (A): three-dimensional structure of Bi2212, indicating the main crystallographic directions. Relevant to our work are the red arrows which show the atomic movements in the in-plane breathing mode (left panel) and those in the out-of-plane buckling mode (right panel). (B): several unit cells of Bi2223 are shown, and the effect of the *c*-axis modulation is visible.

**Fig. 6.** LCO thin film static electron diffraction. Multiple diffraction orders of sharp Bragg spots can be seen indicating a transmission like pattern. These patterns come from transmission through the three dimensional islands observed in AFM measurements. The patterns are indexed based on the tetragonal structure of LCO. (A): the electron beam is incident $45^0$ to in-plane Cu-O bond direction (nodal direction) whereas in (B) the beam is incident along the Cu-O bond direction (antinodal direction). The angle of incidence was around $1.5^0$ in both cases.

**Fig. 7.** Time-resolved diffraction in BSCCO. (A): diffraction intensity change of the (00) rod at different polarizations in optimally doped Bi2212. The laser fluence was 20 mJ/cm$^2$ and the temperature was 50 K. The electron probing was kept along [110] (Fig. 2 B), and $\theta$ is the angle of polarization away from the probing direction (controlled by rotation of a half-wave plate). The dotted lines (and also those in panels (B) to (D)) show the fits to an apparent exponential decay. (B): diffraction intensity change of the (00) rod, from the same sample, obtained with the optical polarization being parallel to the electron probing. By rotating the crystal, the time-dependent change was measured for the two zone axes (Figs. 2 A and 2 B). (C): diffraction intensity change of the (00) rod for an underdoped Bi2212 sample ($T_c$ = 56 K), at two temperatures and two polarizations. (D): diffraction intensity change obtained from a three-layered, optimally doped Bi2223 sample at 45 K for two polarizations.

**Fig. 8**. Scaling of Bragg intensities. Shown are the decay of two distinct Bragg peaks, observed at $s_1$ = 6.3 Å$^{-1}$ (red) and $s_2$ = 4.5 Å$^{-1}$ (blue). The green curve is obtained by multiplying the data at $s = s_2$ by the factor of $(s_1/s_2)^2$, according to Eq. (2), and its match with the data at $s = s_1$ confirms the structurally induced diffraction changes following the carrier excitation.

**Fig. 9.** Experimental and theoretical intensity transients. (A): lattice temperature derived from diffraction using Eqs. (1) and (3), for different polarizations, along [010] (blue dots) and [110] (red dots), in optimally doped Bi2212. From the three-temperature model described in the text, we obtain $\lambda$ = 0.08 for $\vec{E}$ //[110] ($T_l$, red solid line) and $\lambda$=0.55 for $\vec{E}$ //[010] ($T_l$, blue solid line). The electronic ($T_e$, dashed lines) and hot-phonon ($T_p$, solid lines) temperatures are also displayed. (B): derivatives of the (00) diffraction intensity

derived from Fig. 2A for different polarizations. (C): derivatives of the simulated lattice temperature within the three-temperature model, for different $\lambda$ with a fixed anharmonic coupling time $\tau_a$ = 2.8 ps [also shown in (B)] and for different $\tau_a$ with a fixed $\lambda$ = 0.26 (*inset*). The clear shift of the minimum position is only observed when $\lambda$ is varied (black dotted lines). (D): three-temperature model analysis for underdoped Bi2212. (E): three-temperature model analysis for Bi2223.

**Fig. 10.** Temperature dependence and phase diagram. (A): fluence dependence of the intensity decay along the (010) direction of optimally doped Bi2212. In the inset, the fluence dependence of the overall time constant of the decay, obtained from a simple exponential fit (filled black symbols) is shown together with the time constant associated to the electron-phonon coupling time (empty black symbols), obtained from Fig. 10 B. (B): fluence dependence of the derivative of the intensity decay along (010) in optimally doped Bi2212. The characteristic time related to the electron-phonon coupling process is estimated from the minimum in these curves (open black symbols in the inset of Fig. 10 A). (C): lattice-equilibrium temperature dependence of the time constant for both polarizations in all samples. (D): the doping dependence of the coupling constant ($\lambda$) along the [010] and [110] directions in Bi2212 (blue and red dots, respectively) and its anisotropy ($\Delta\lambda$ between the two directions; black solid line), as well as $\lambda$ along the [010] and [110] directions in Bi2223 (green and orange dots, respectively) and the extrapolated anisotropy (black dashed line). A qualitative sketch of the upper critical field and Cooper-pair coherence length (green and violet lines) is also shown.

**Fig. 11.** Excitation scheme and Fermi Surface. (A): possible excitation scheme of different Pomeranchuk modes. (B): Fermi surface predicted for cuprates. The cross–like blurred feature is the FS of the stripes, while the nodal arcs are the FS of nodal particles observed in ARPES. (C): possible Pomeranchuk modes excited by the laser for the electric field in the antinodal direction and (D): the electric field in the nodal direction. The blue line is the equilibrium Fermi surface while the red lines is the out of equilibrium Fermi surface after the laser excitation.

**Fig. 12.** BSCCO and LCO optical spectra and temperature dependence. (A): the difference spectrum $\sigma_1(T = 280)- \sigma_1(T = 20)$ for Bi2223 is displayed, the low energy region is magnified in the inset. (B): the optical conductivity of Bi2223 at different temperatures is shown together with a Drude-Lorentz decomposition of the spectrum, the low energy region is magnified in the inset. (C): the optical conductivity of different cuprates is shown for comparison, together with the theoretical spectrum obtained in LCO by the charge-stripes model.

**Fig. 13.** Theoretical description of the optical excitation. (A): symmetry of charge excitation for 1.2 eV light polarized along the direction of the charge stripes (*y*, see Fig. 13 B). (B): the charge distribution inferred by the stripes model in LCO. (C): symmetry of charge excitation for 1.2 eV light polarized perpendicular to the direction of the charge stripes (*x*, see Fig. 13 B). (D): theoretically derived optical conductivity spectrum in LCO.

**Fig. 14.** LCO dynamical diffraction pattern**.** The effect of laser excitation on the diffraction pattern is shown by displaying diffraction difference frames at different times after the arrival of the laser pulse. These frames were obtained by referencing diffraction patterns at each time delay to a negative-time frame by subtraction. Bright regions indicate positive intensity and dark regions show the negative intensity. Laser excitation results in a movement of the Bragg spot profiles along the *c*-axis direction to lower momentum transfer values. The laser induced changes reach a maximum around ~120 ps and relax back to equilibrium on longer time scale (1 ns). These patterns were obtained at room temperature for a laser fluence of 20.6 mJ/cm$^2$

**Fig. 15.** Bragg peak position shift and scaling of different orders**.** Detailed analysis of the photo-induced changes of the Bragg peaks for the pattern shown in Fig. 6**.** (A): the profiles of the 008 and 0010 Bragg spots are shown at selected time delays. The profiles for time delays between 0 and 112 ps are displaced vertically for clarity. The time delays for the curves displayed in the top panel are t = −233, −33, −4, 8, 20, 32, 44, 56, 68, 80, 92 and 112 ps. For the curves in the bottom panel t = 112, 147, 217, 317, 617, 917 and 1217 ps. For the curves in the top panel, a structural isosbestic point can be seen for both Bragg spots (see text). The curves in the bottom do not cross, the center positions of the Bragg spots shift back to equilibrium continuously. (B): the relative changes of the center positions of 0010 (blue), 008 (red) Bragg spots and direct electron beam (black) are plotted as a function of time. The center positions are obtained by fitting the curves in (A) to a single Gaussian form. The fit to such function is good except for the region shown by a transparent yellow strip during which a single Gaussian can not describe the profiles

since at least two structural phases coexists. We also plot the shift of the 0010 Bragg spot (blue curve) scaled by % 80 as the dashed blue curve. The agreement between this and the red curve is consistent with a true structural change in which the shifts of the Bragg peaks scale proportionally to the order numbers since $\Delta s/s = -\Delta c/c$. Furthermore, no substantial movement of the direct beam is observed, indicating that the observed effects can not be coming from shifting of the entire pattern.

**Fig. 16. Structural isosbestic point and fluence dependence.** The Bragg spot profiles of the 0010 spot is shown at different delay times for five laser fluences (2.8, 7.0, 11.3, 14.3 and 20.6 mJ / cm$^2$) at room temperature. The amplitude of the profiles is normalized by the height of the highest curve at each intensity and curves at different fluences are displaced vertically for clarity. No laser induced change can be observed at the lowest fluence (2.8 mJ/cm$^2$)**.** Above this fluence, transitions into a higher c axis constant state can be observed as evidenced by the existence of structural isosbestic points. The change in the center position of the Bragg spot ($\Delta$s) increases linearly with increasing intensity. (A): phase transition region showing the times t = −233, −33, −4, 8, 20, 32, 44, 56, 68, 80, 92 and 112 ps. (B): relaxation region showing the times t = 112, 147, 217, 317, 617, 917 and 1217 ps. Here, no crossing point is observed at any fluence, all the peaks relax back to equilibrium by continuously shifting their center positions.

**Fig. 17. Fluence dependence of the expansion.** (A): the change in the *c*-axis constant ($\Delta$c) is plotted as a function of time for each laser fluence (2.8 (cyan), 7.0 (blue), 11.3 (green), 14.3 (red) and 20.6 mJ / cm$^2$ (black)). The *c*-axis lattice constant at each time delay was obtained by fitting the profiles in Fig. 16 to a single Gaussian form. The inset shows the same curves normalized. $\Delta c$ relaxes back to ground state with a time constant of about 300 ps independent on the laser fluence. Even after 1 ns, the system does not completely relax back to equilibrium; a small residual expansion is present and has a much longer time constant. (B): $\Delta c$ (130 ps) around the dotted line shown in (A) was measured as a function of laser fluence at both 20 K and 300 K. Below a threshold intensity of about ~ 5 mJ/cm$^2$, no change can be observed. Above this threshold, $\Delta c$ grows linearly with the laser fluence. The top horizontal scale shows the number of photons absorbed per copper site calculated using the absorption coefficient (see text). The threshold fluence corresponds to ~0.1 photons absorbed per copper site. The red line is a linear fit to room temperature data above the threshold value.

**Fig. 18.** Cohesion energy calculations. (A): an elementary unit of La$_2$CuO$_4$ in the tetragonal structure that includes a CuO$_6$ octahedron and two La atoms is shown. The oxygen atoms in the CuO$_2$ planes are labeled as O(2) and out of plane oxygen atoms are labeled as O(1). O(2) atoms are counted as ½ in the summations, since they are shared by two units. La$_2$CuO$_4$ in the tetragonal structure can be described by four structural parameters; distance between CuO$_2$ planes (c/2), Cu – O(1) distance ($d_1$), Cu – O(2) distance ($d_2$) and Cu – La distance ($d_3$). In the excited state, the valance of a Cu atom is decreased by $\delta_p$ and the valance of an O(2) atom is increased by $\delta_p$ /2, where $\delta_p$ is the number of photons absorbed per copper atom. (B): calculated cohesion energy as a function of the *c*-axis constant in the ground ($\delta_p=0$) and excited states with $\delta_p=0.3$ assuming a uniform expansion along the *c*-axis direction. Calculated cohesion energy in the excited state is accurate up to an overall constant. (C): comparison of the calculated Δc using different methods with the experiment (blue points). The dashed green line is the result of the calculation without considering the van der Walls interactions, dashed red and black lines are the calculated Δc after taking van der Walls interaction into account and assuming uniform expansion or independently variable parameters respectively (see text). All of the calculations reproduce the magnitude of the observed expansion remarkably well.

**Fig. 19.** Energy landscape. A depiction of the excited state energy landscape is shown.

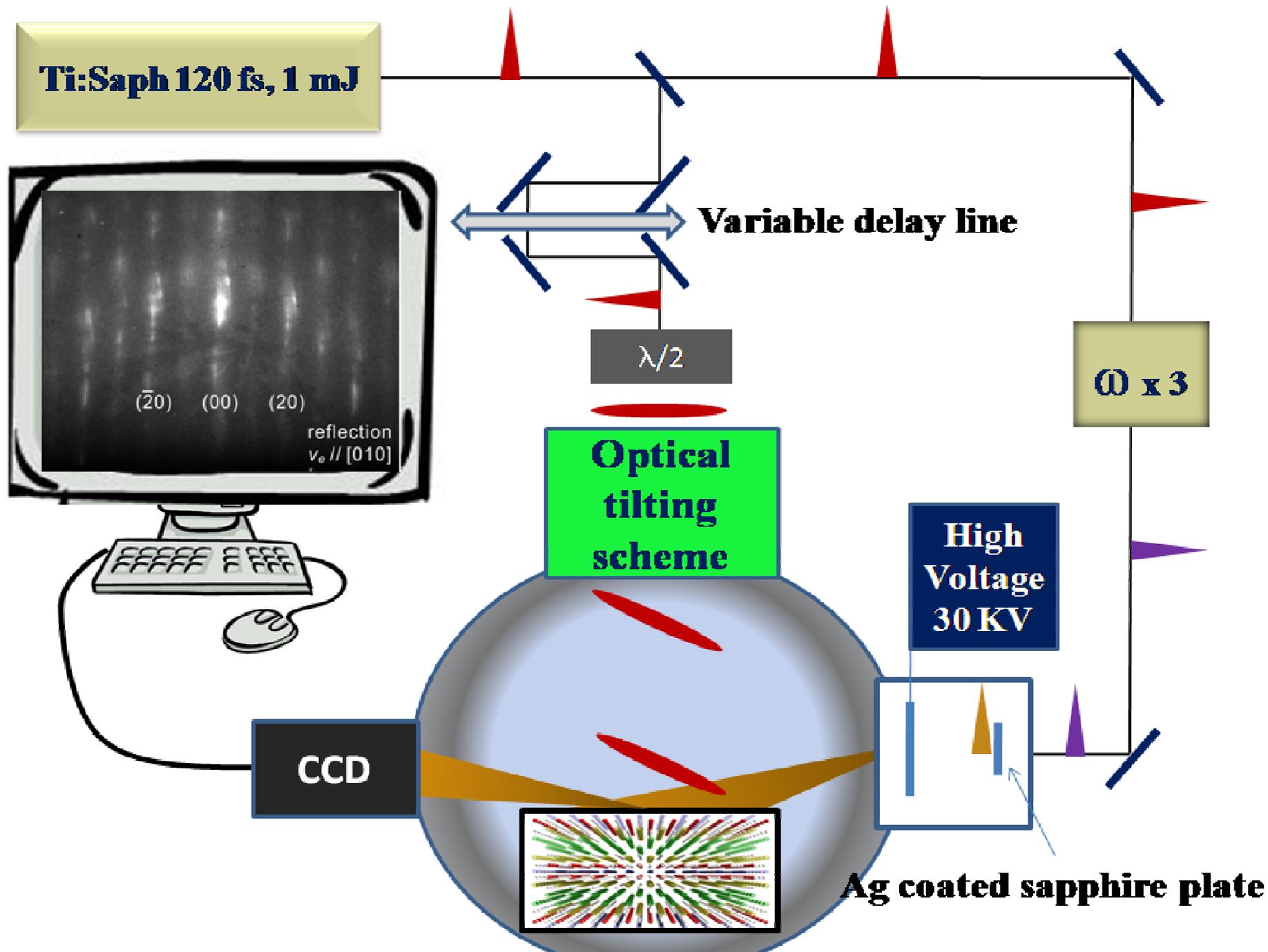

Figure 1: Experimental Set-up

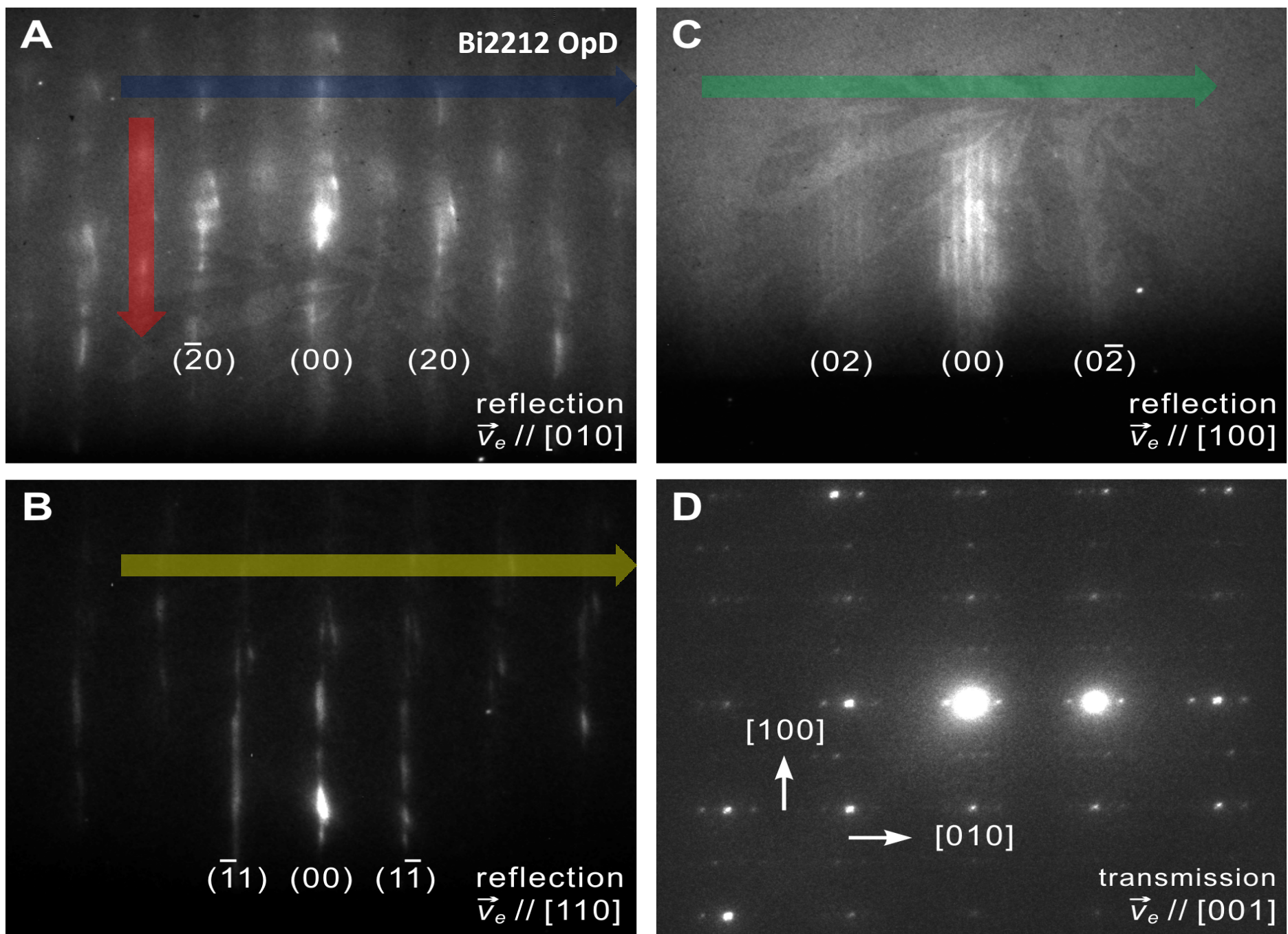

Figure 2: OpD Bi2212 static diffraction pattern

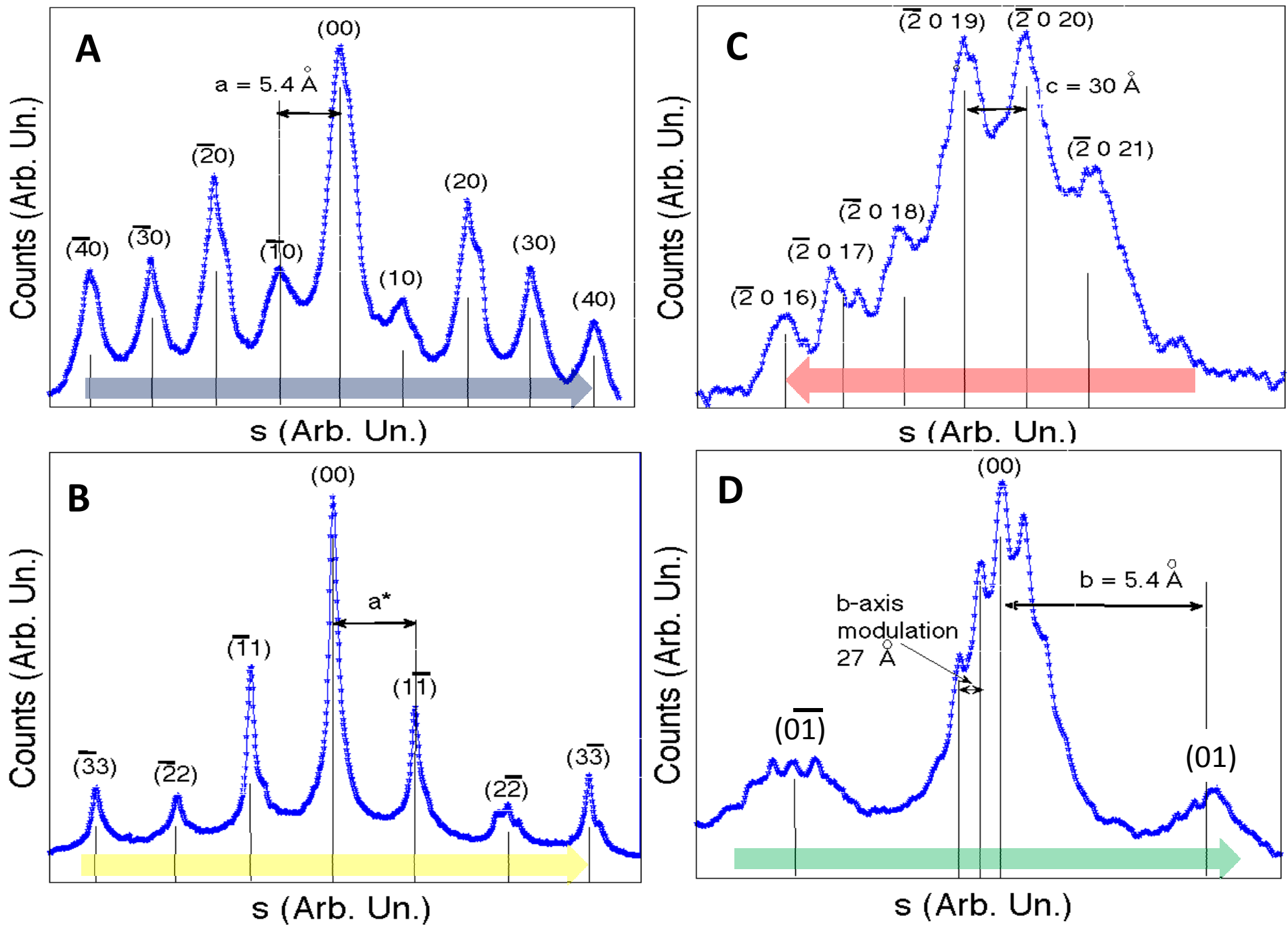

Figure 3: OpD Bi2212 Bragg peaks

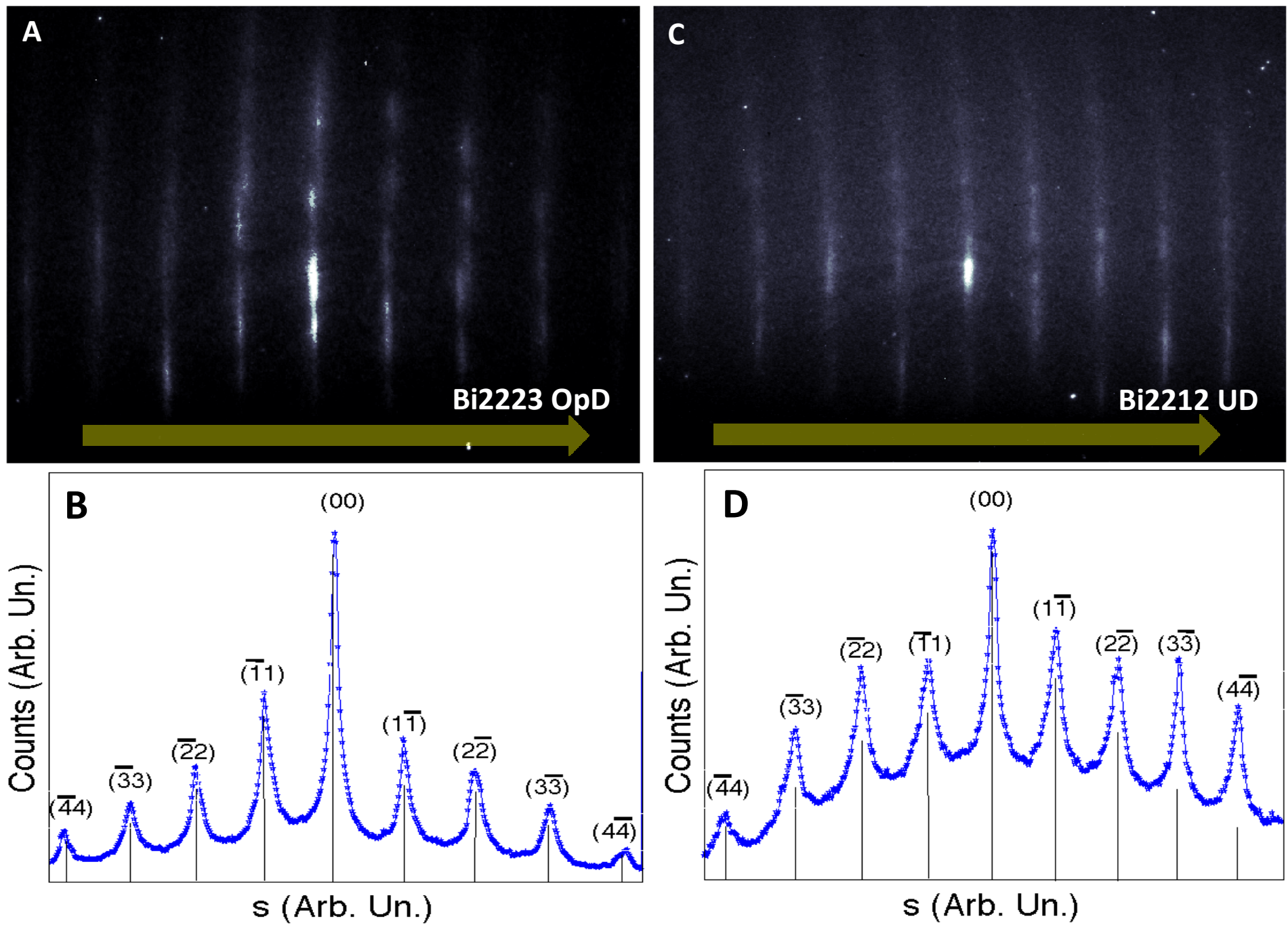

Figure 4: OpD Bi2223 and UD Bi2212 diffraction patterns and Bragg peaks

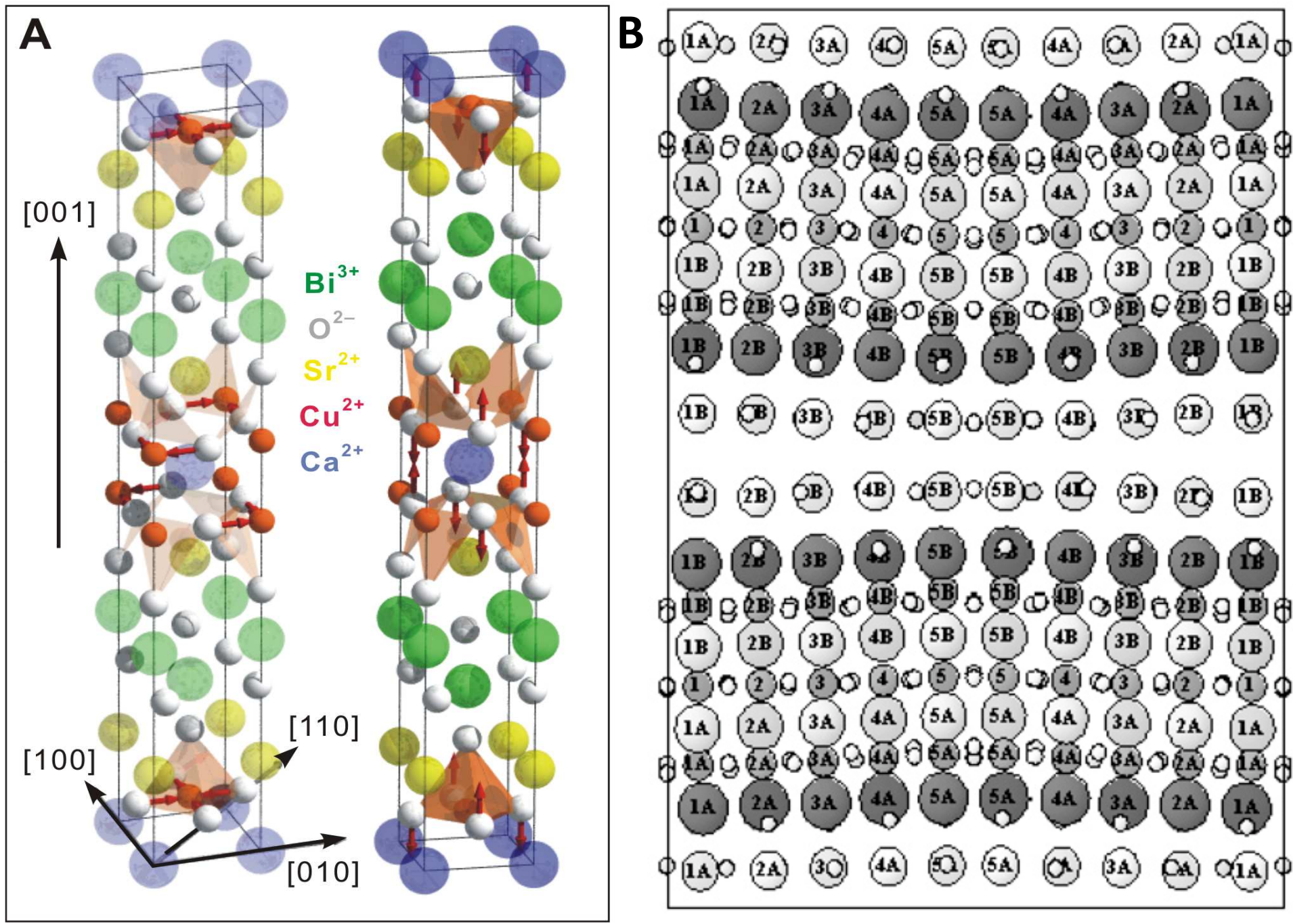

Figure 5: Bi2212 unit cell and Bi2223 modulated supercell

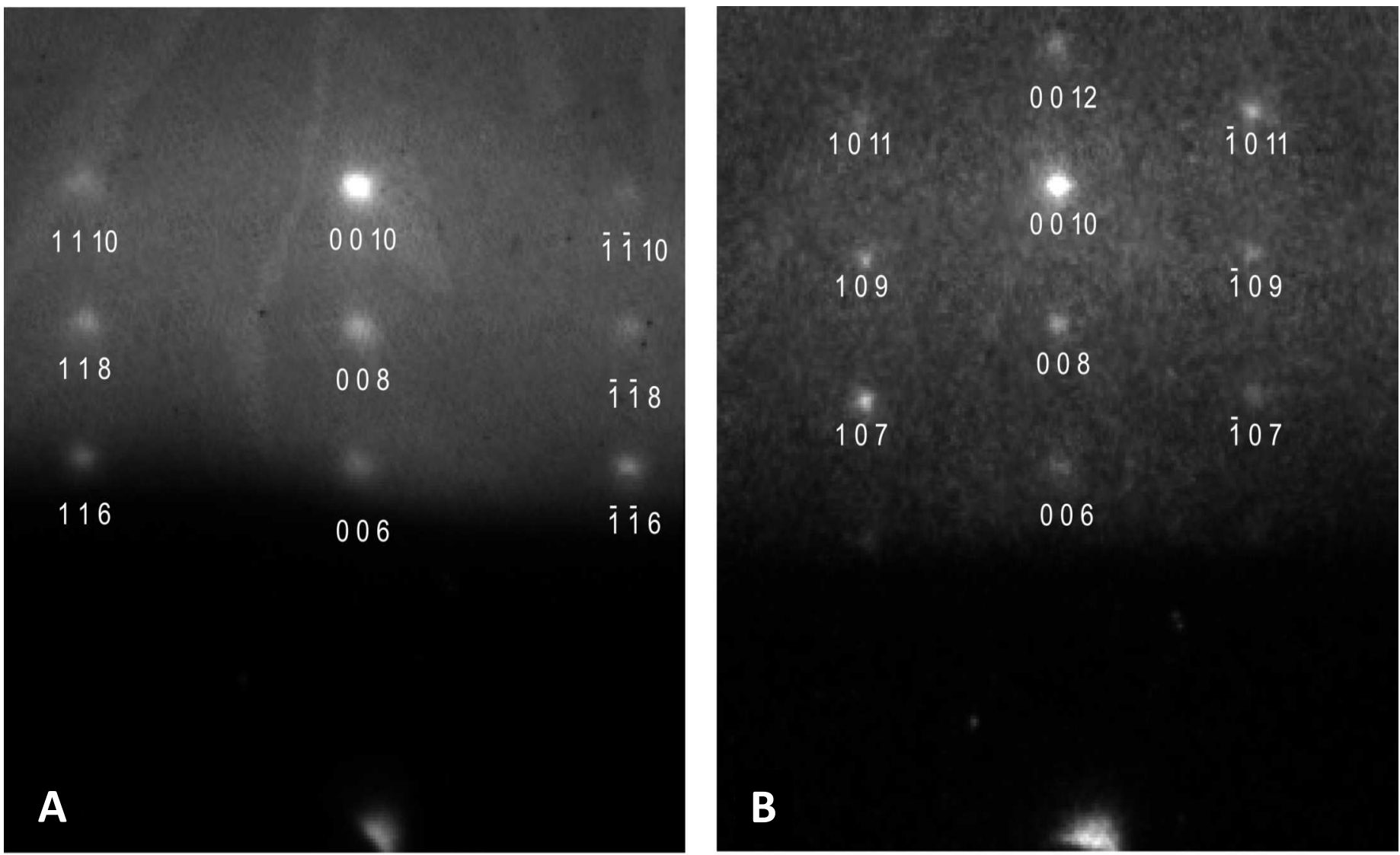

Figure 6: LCO thin film static electron diffraction

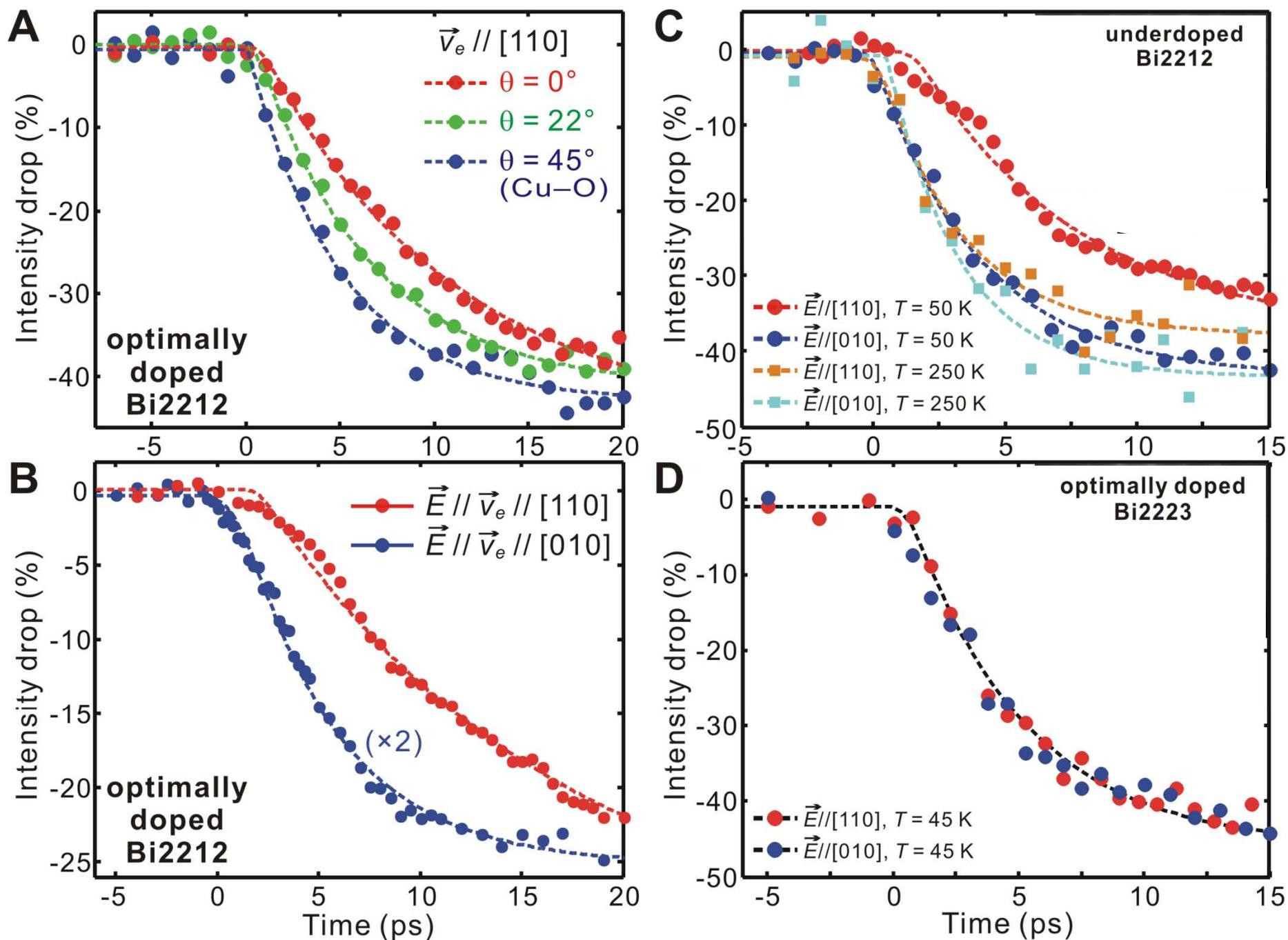

Figure 7: Time-resolved diffraction in BSCCO.

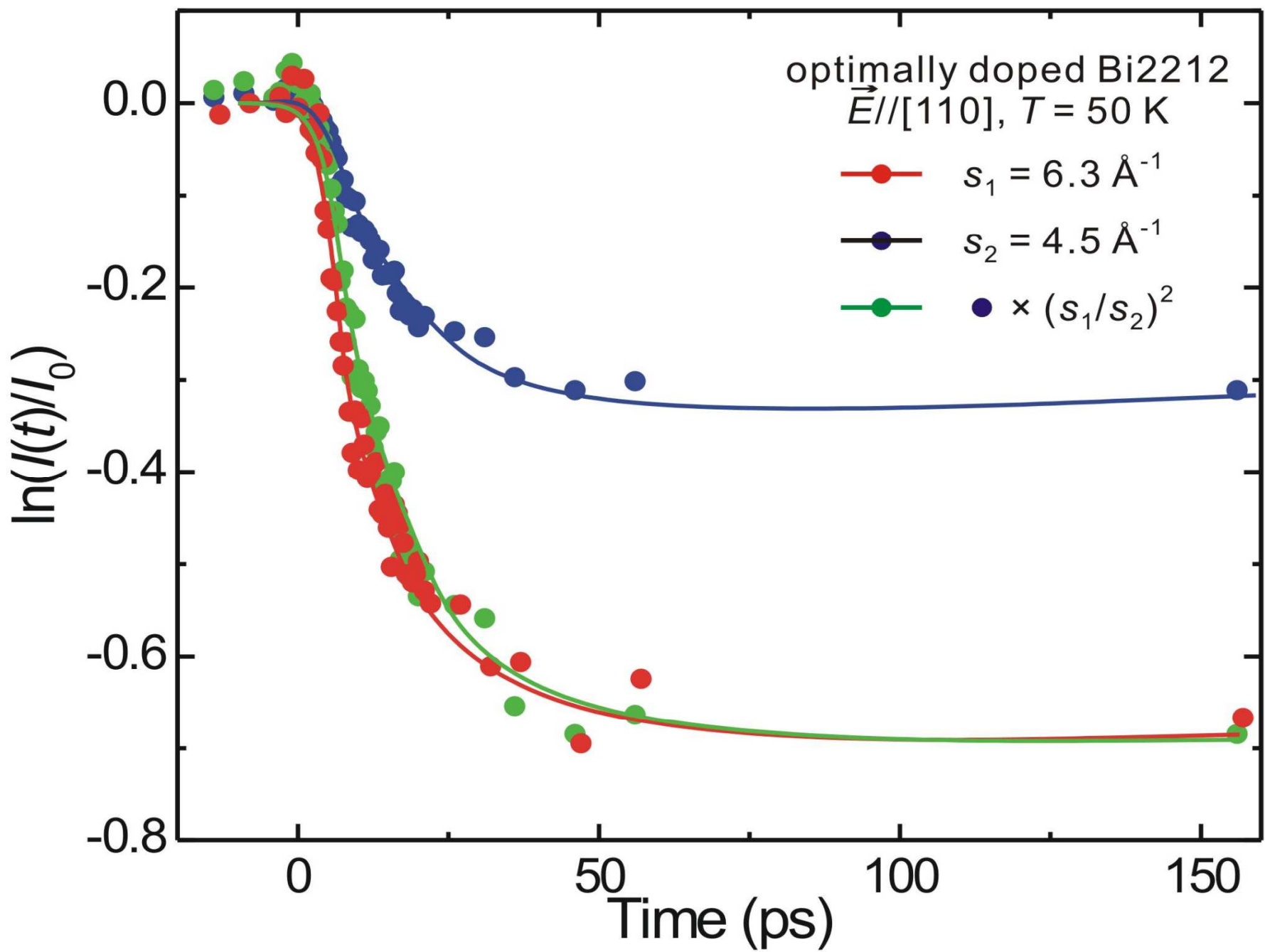

Figure 8: Scaling of Bragg intensities

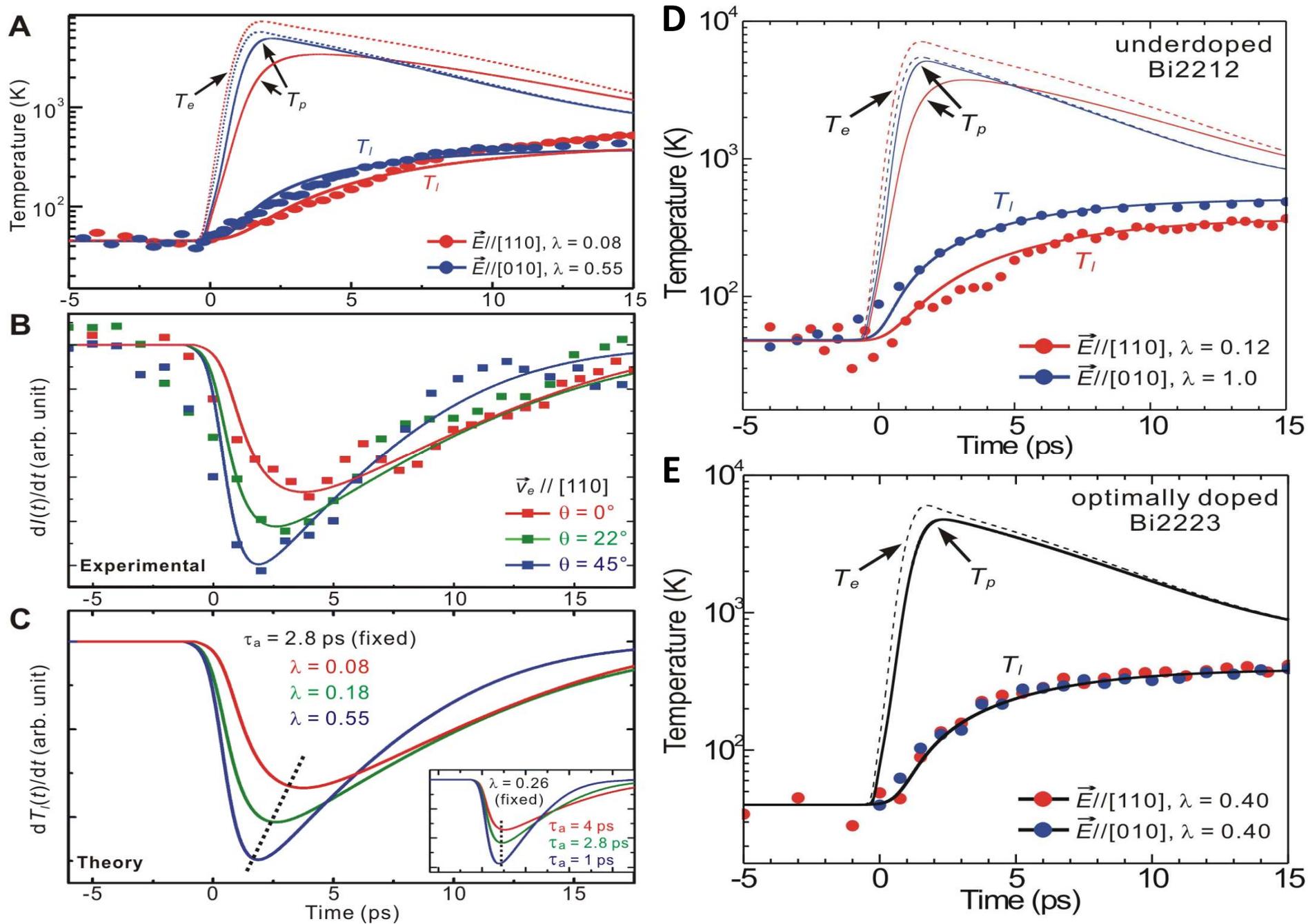

Figure 9: Experimental and theoretical intensity transients

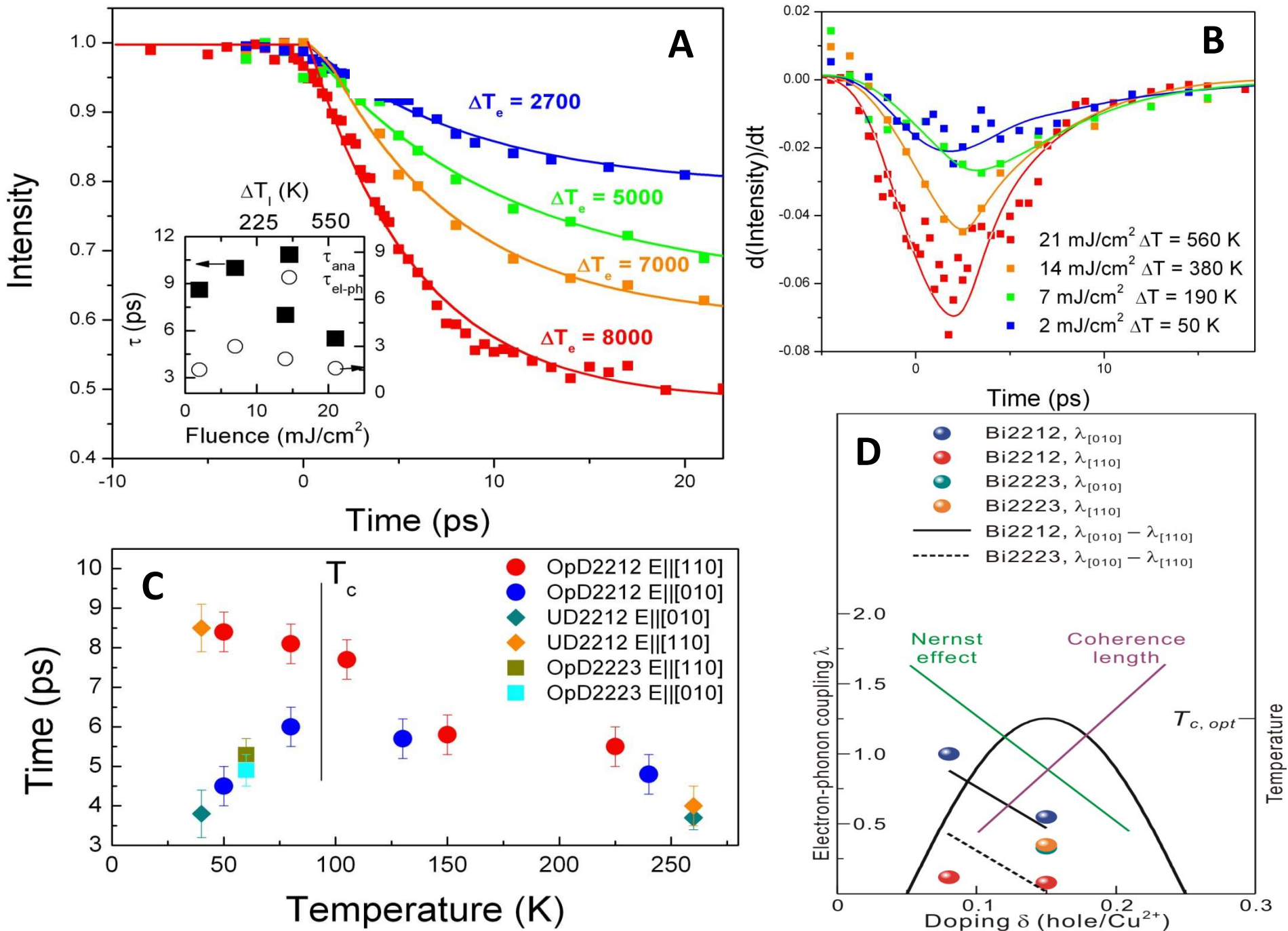

Figure 10: Temperature dependence and phase diagram

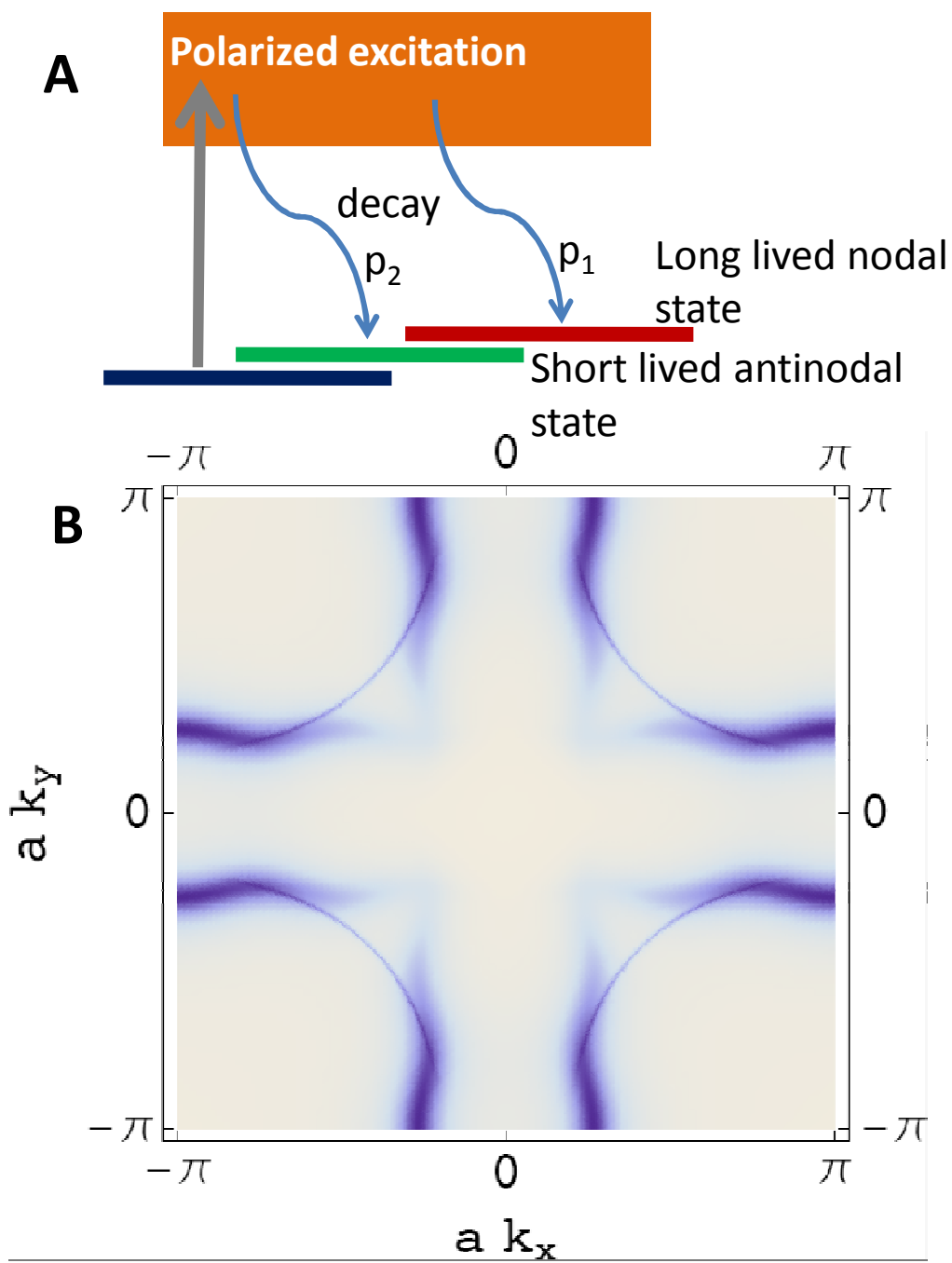
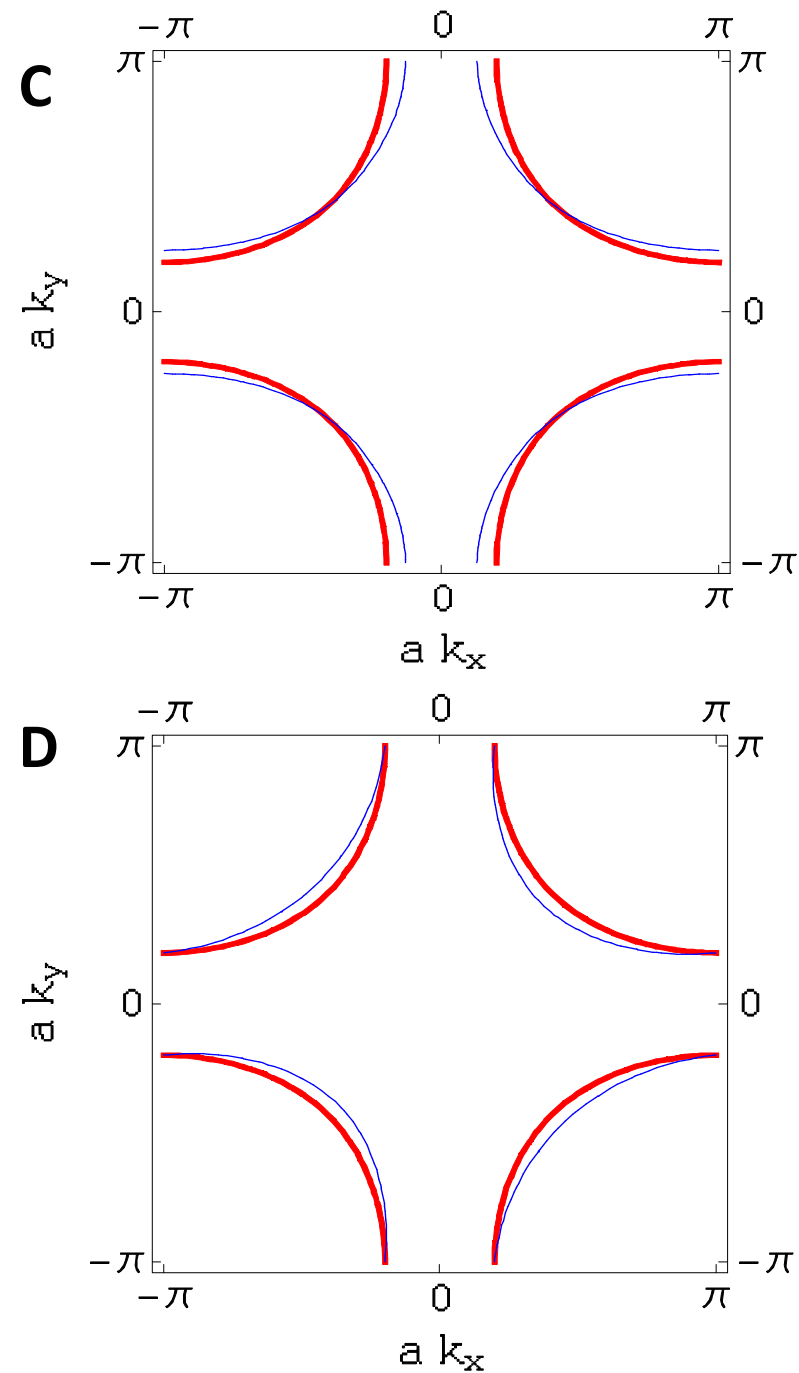

Figure 11: Excitation scheme and Fermi surface

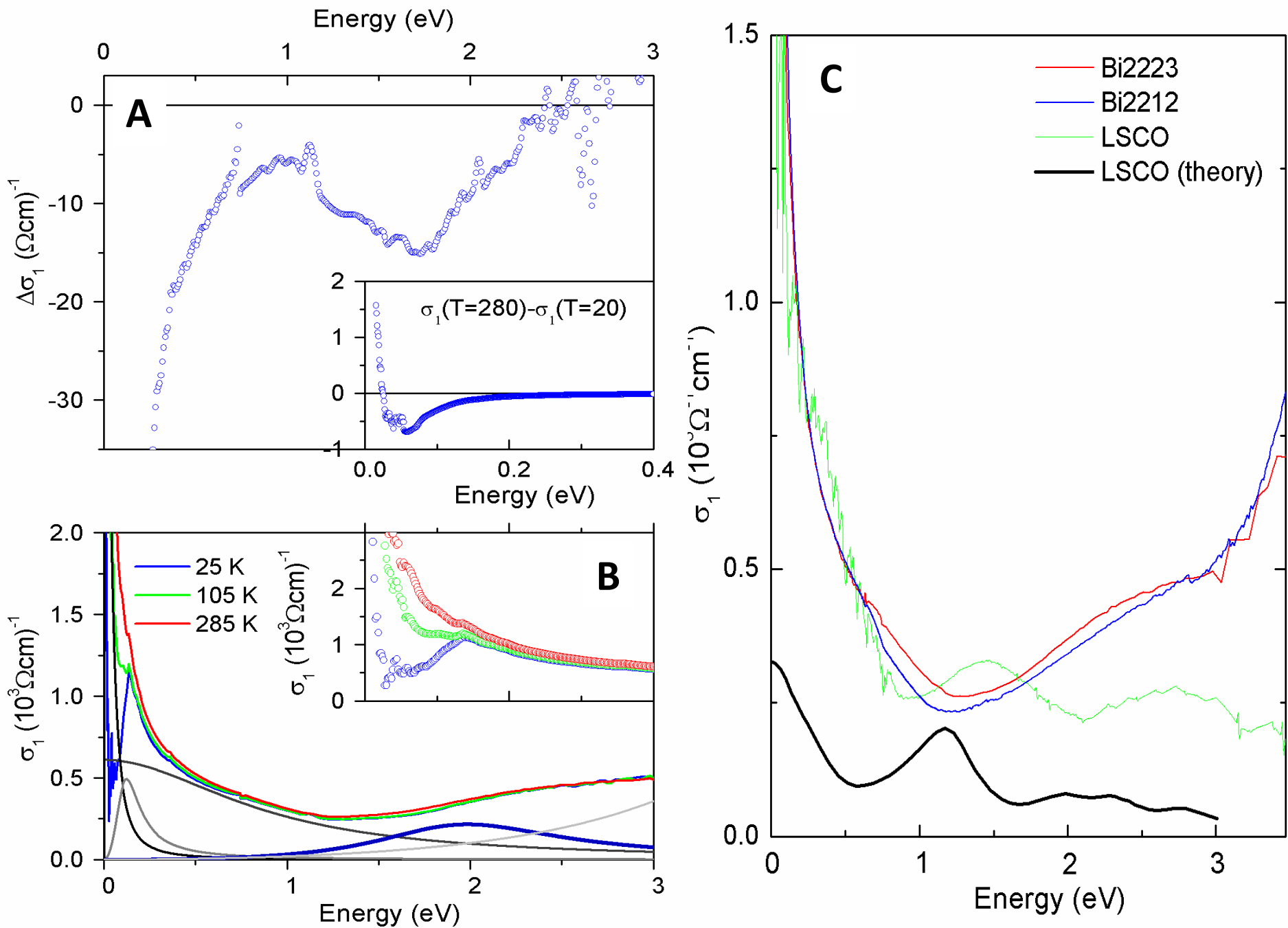

Figure 12: BSCCO and LSCO optical spectra and temperature dependence

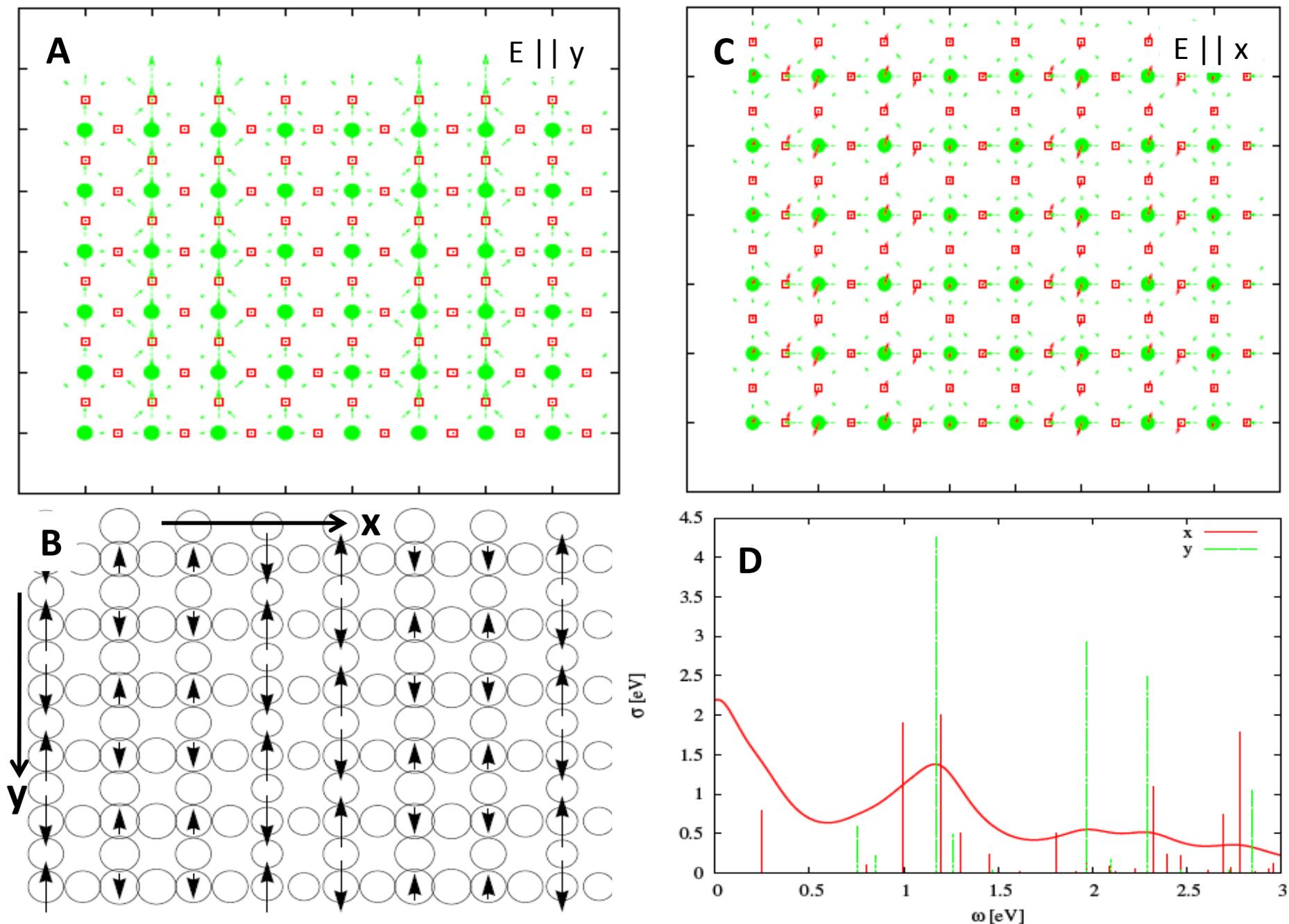

Figure 13: Theoretical description of the optical excitation

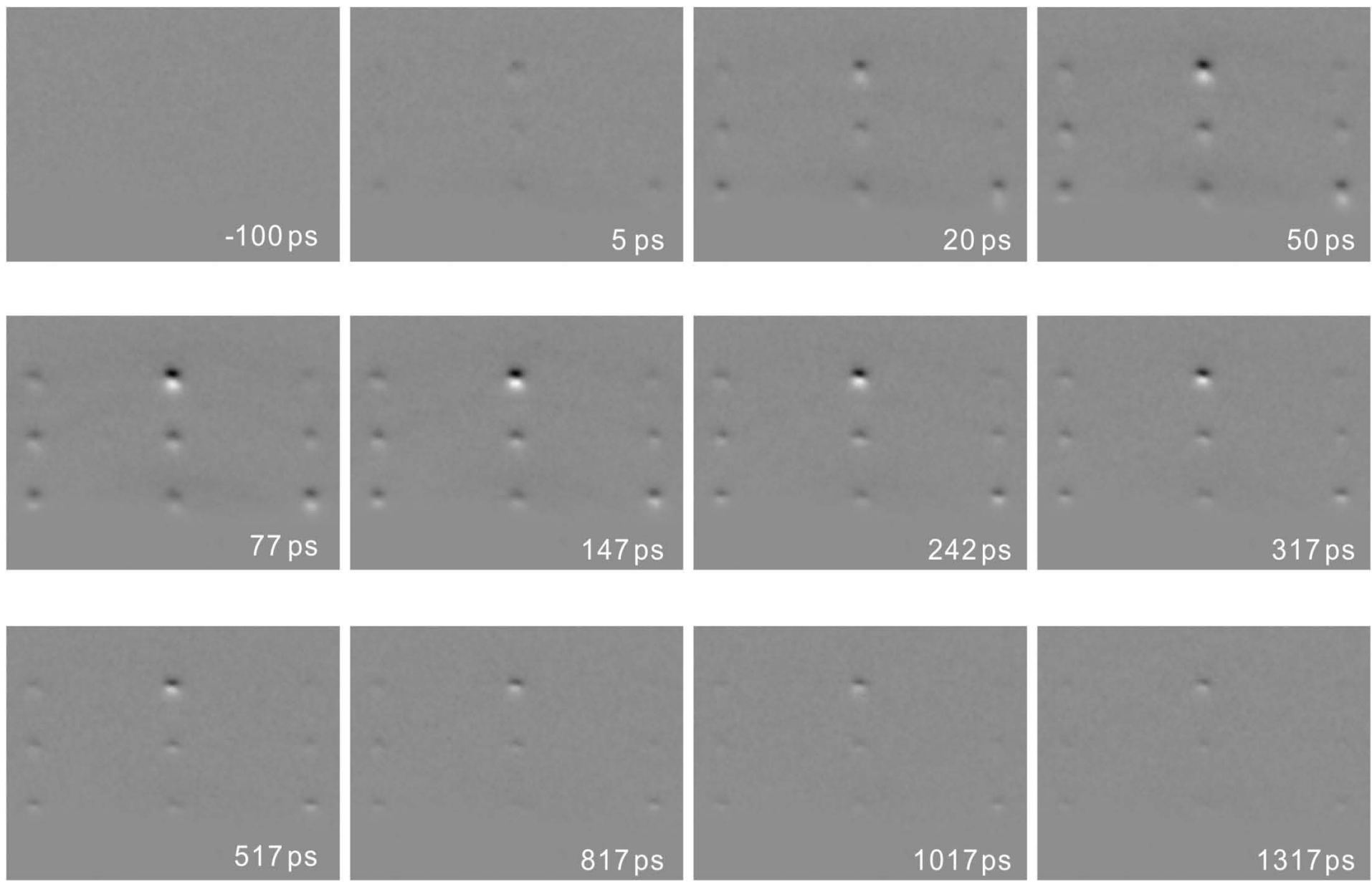

Figure 14: LCO dynamical diffraction pattern

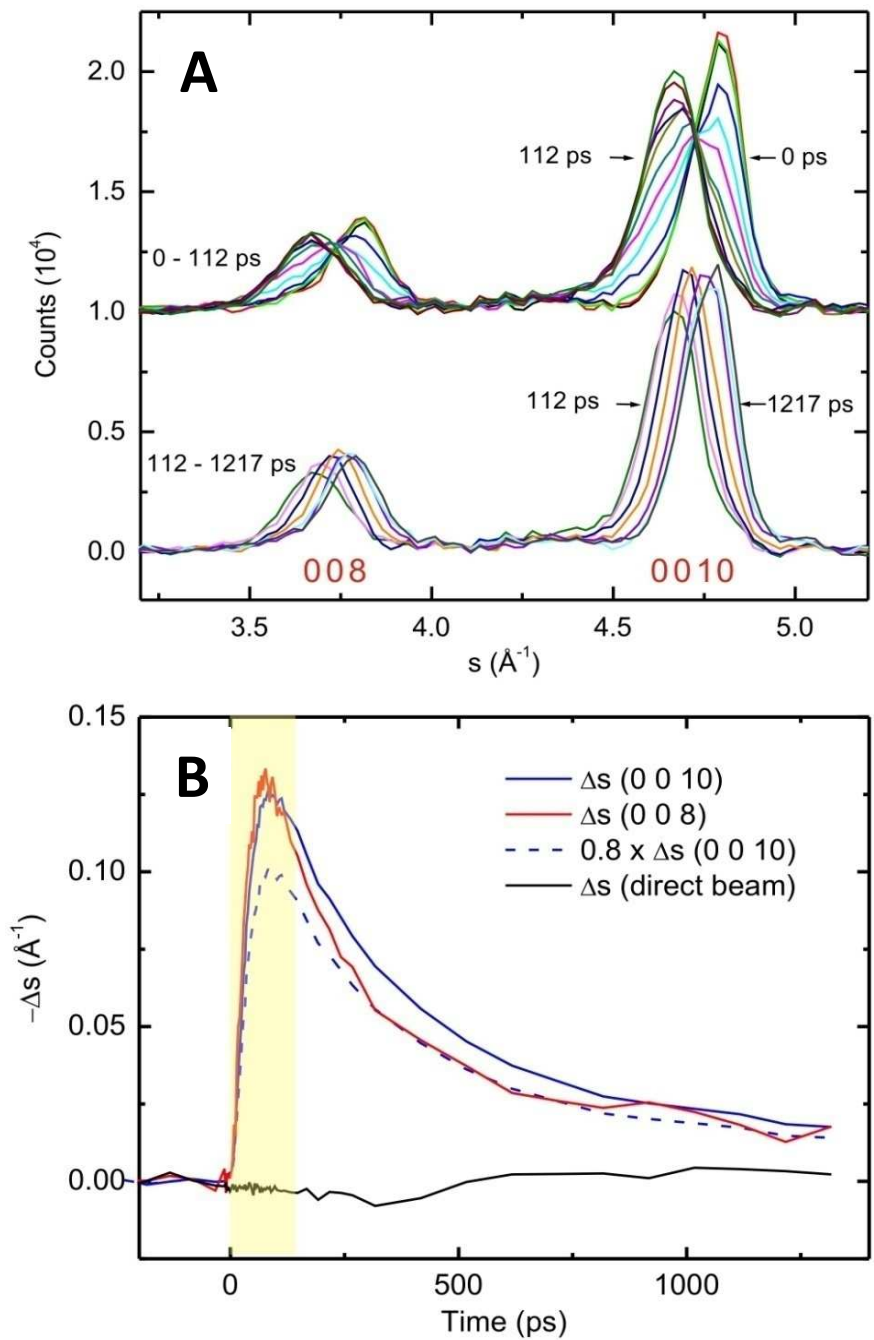

Figure 15: Bragg peak position shift and scaling of different orders

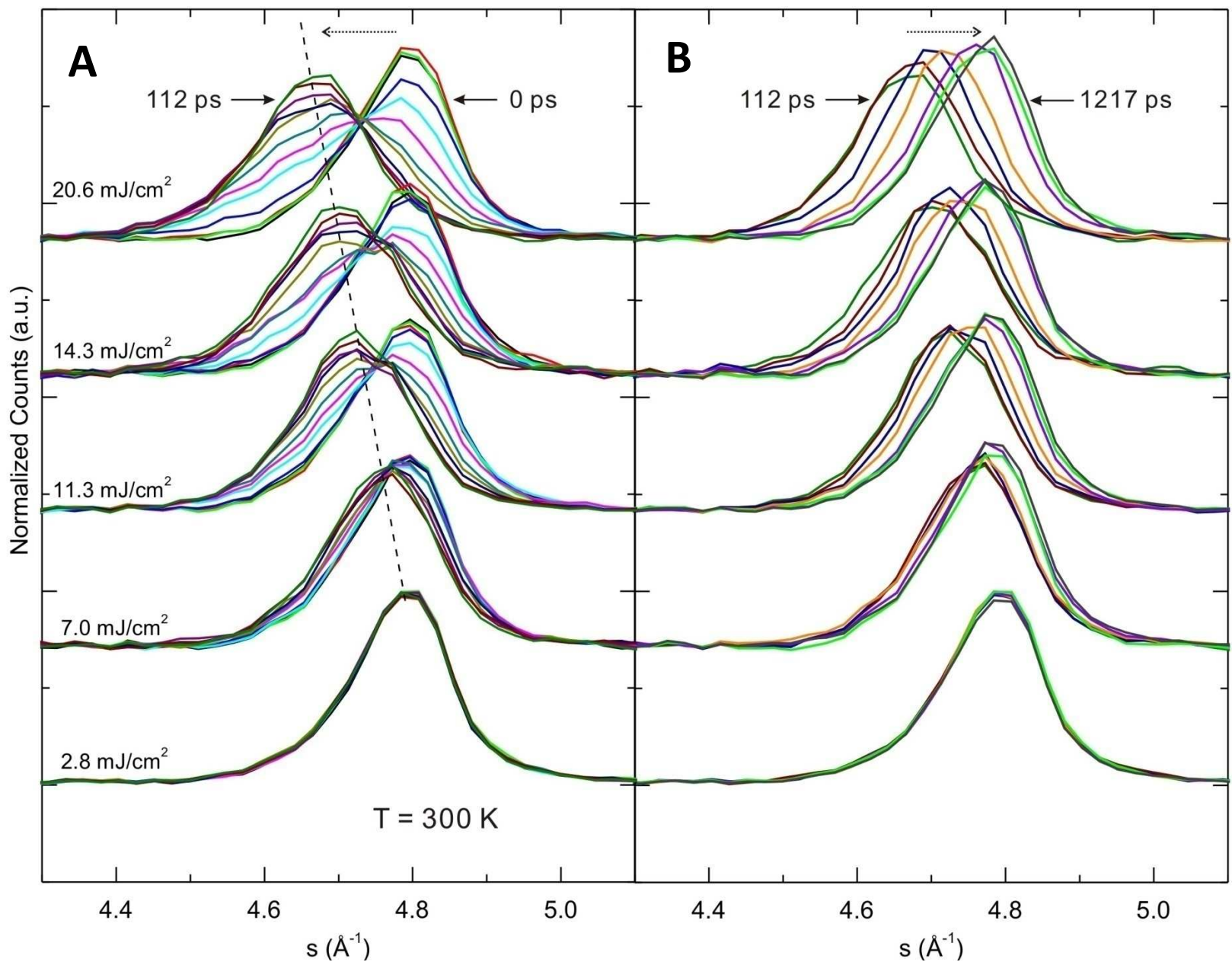

Figure 16: Structural isosbestic point and fluence dependence

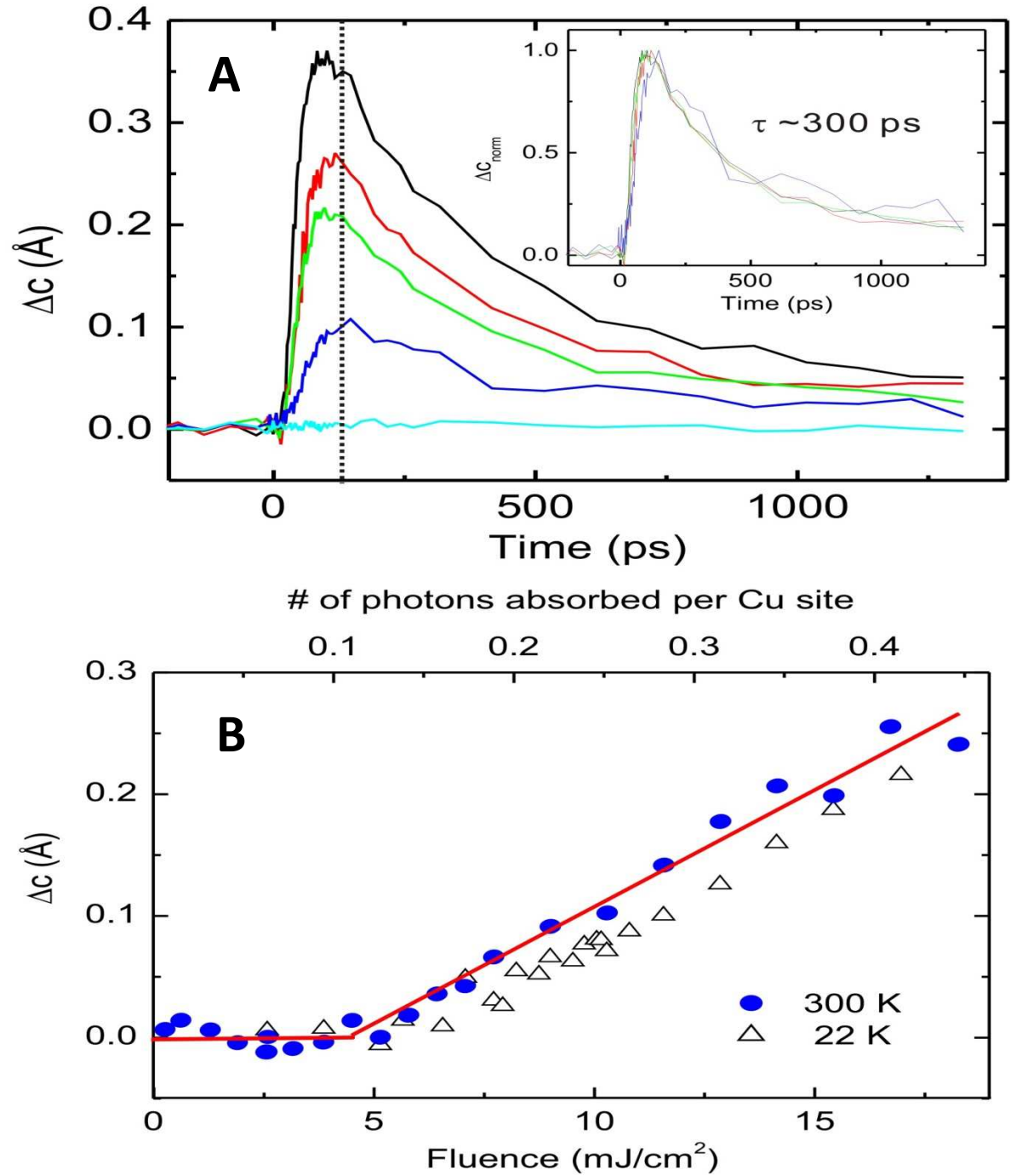

Figure 17: Fluence dependence of the expansion

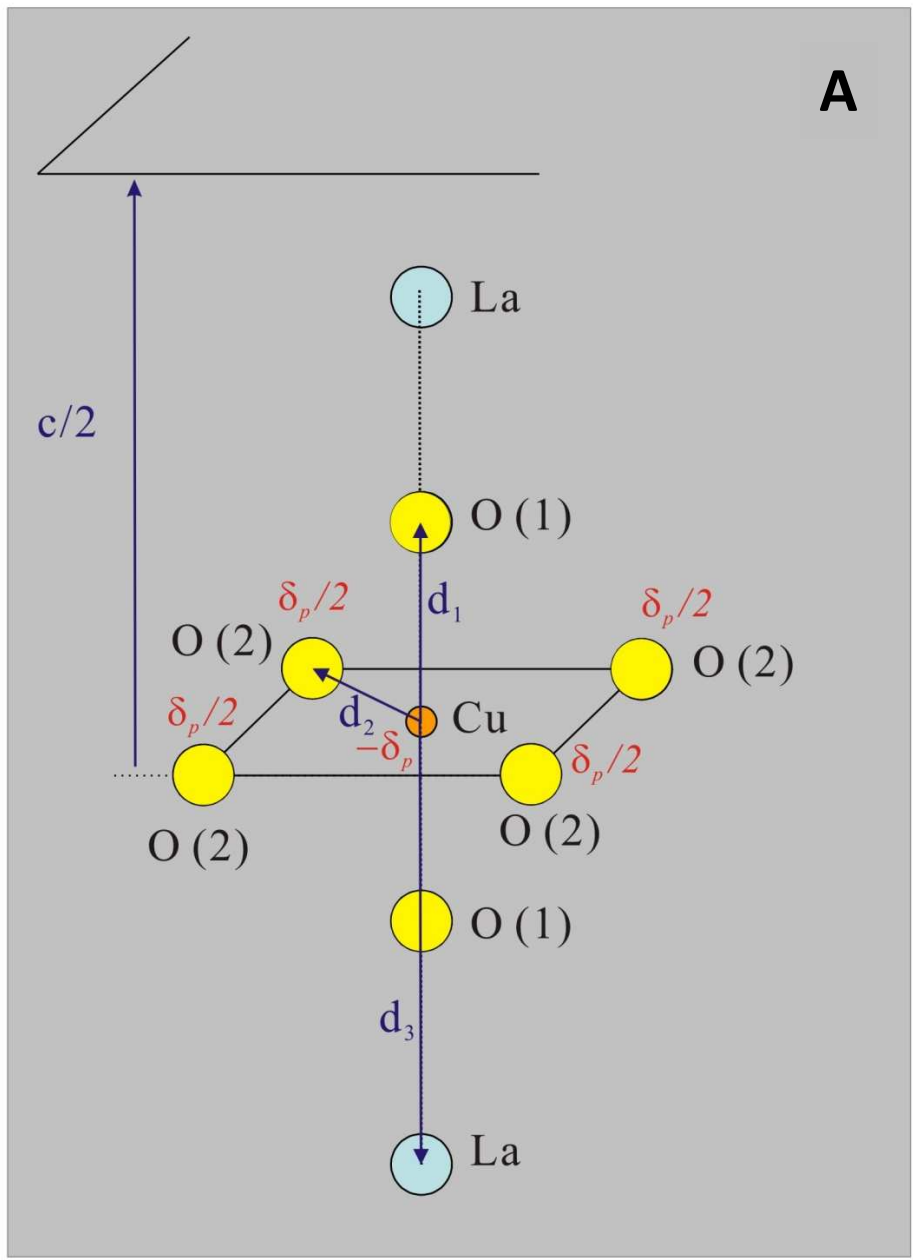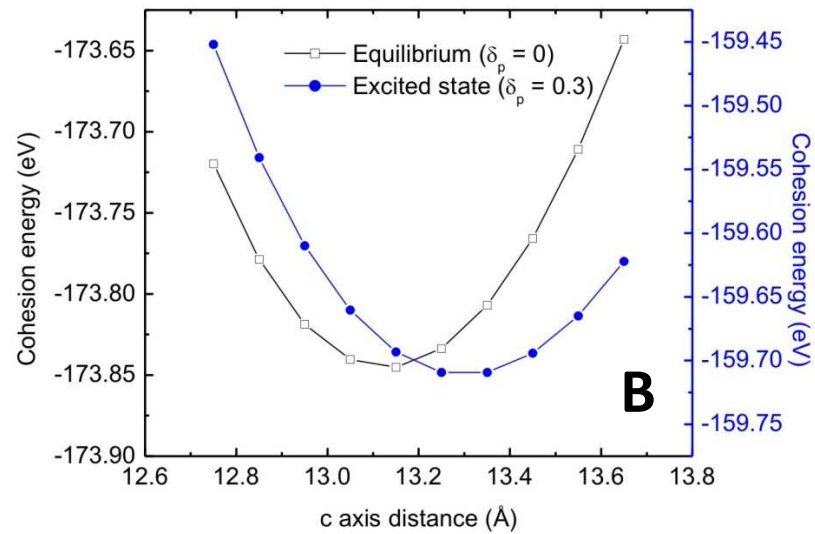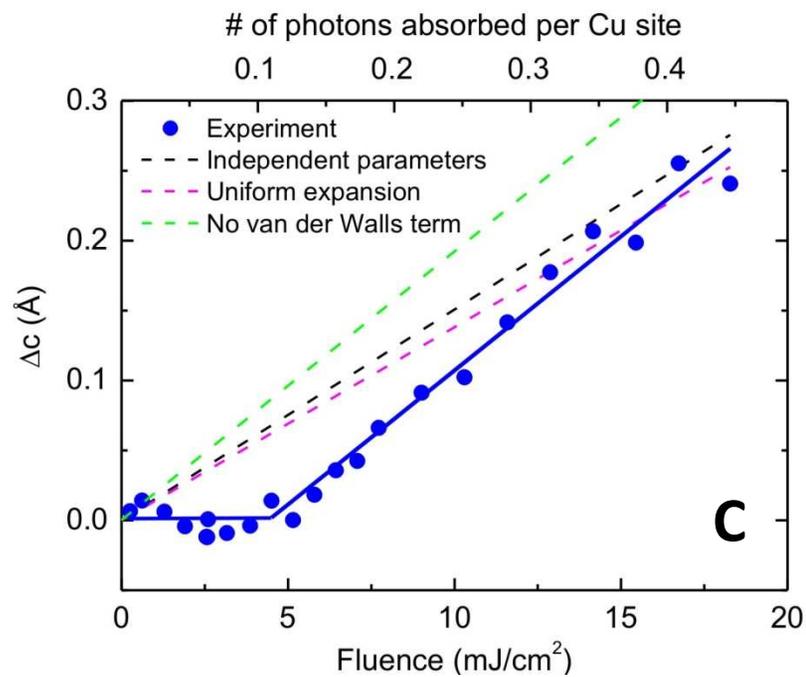

Figure 18: Cohesion energy calculations

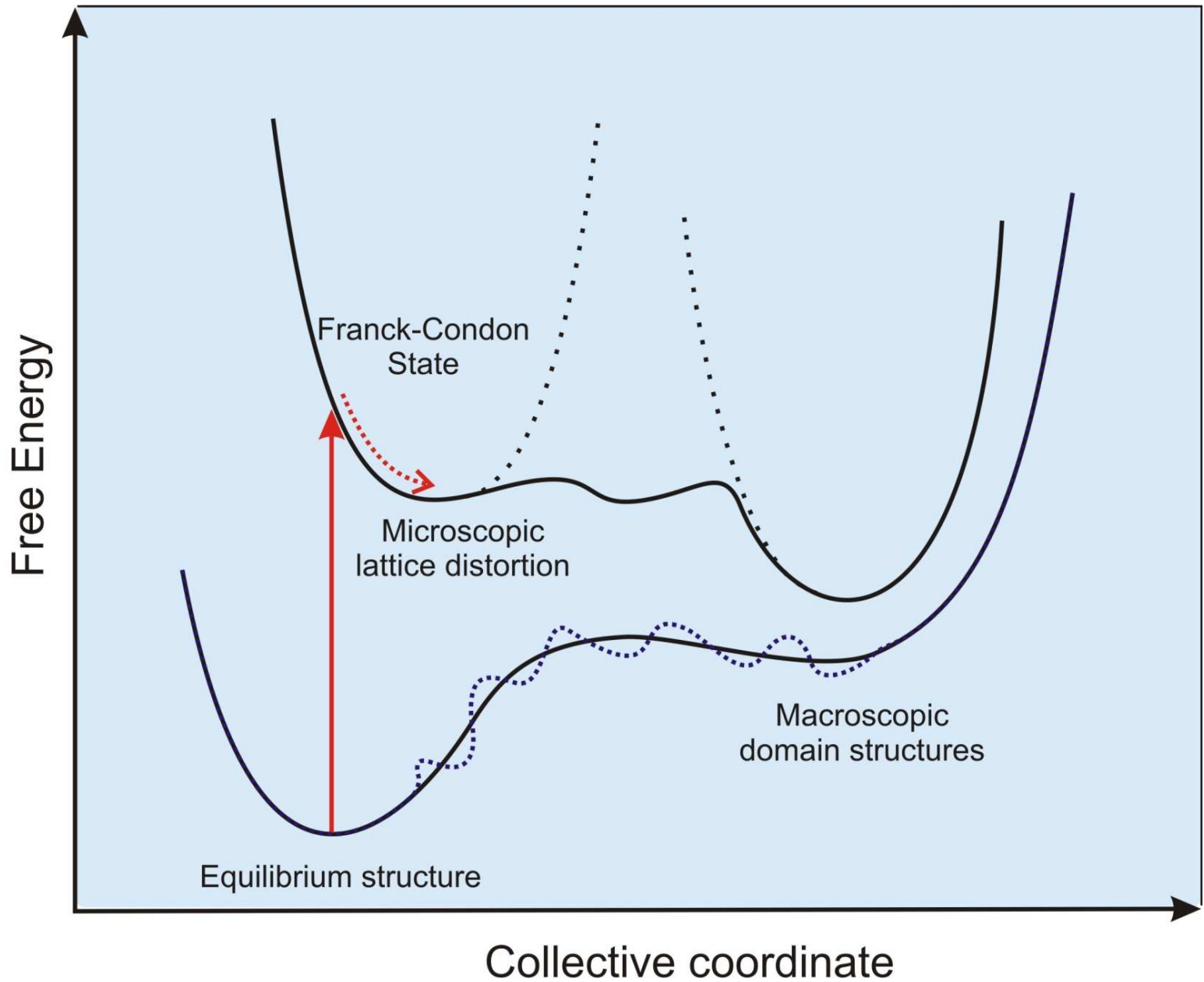

Figure 19: Energy landscape